\documentclass[twocolumn]{aastex62}
\usepackage{amsmath}
\usepackage{enumerate}
\usepackage{hyperref}
\usepackage{color}
\usepackage{subfigure}
\usepackage{color}
\usepackage{longtable}

\received{}
\revised{}
\accepted{}
\submitjournal{}

\shorttitle{Molecular gas in galaxy pairs}
\shortauthors{Pan et al.}

\begin{document}

\title{The Effect of Galaxy Interactions on Molecular Gas Properties}

\email{hapan@asiaa.sinica.edu.tw}

\author{Hsi-An Pan}
\affil{Institute of Astronomy and Astrophysics, Academia Sinica, AS/NTU Astronomy-Mathematics Building, No.1, Sec. 4, Roosevelt Rd, Taipei 10617, Taiwan}

\author{Lihwai Lin}
\affiliation{Institute of Astronomy and Astrophysics, Academia Sinica, AS/NTU Astronomy-Mathematics Building, No.1, Sec. 4, Roosevelt Rd, Taipei 10617, Taiwan}

\author{Bau-Ching Hsieh}
\affiliation{Institute of Astronomy and Astrophysics, Academia Sinica, AS/NTU Astronomy-Mathematics Building, No.1, Sec. 4, Roosevelt Rd, Taipei 10617, Taiwan}

\author{Ting Xiao}
\affiliation{Department of Physics, Zhejiang University, Hangzhou 310027, China}
\affiliation{Shanghai Astronomical Observatory, Nandan Road 80, Shanghai 200030, China}

\author{Yang Gao}
\affiliation{Shanghai Astronomical Observatory, Nandan Road 80, Shanghai 200030, China}

\author{Sara L. Ellison}
\affiliation{Department of Physics and Astronomy, University of Victoria, Finnerty Road, Victoria, British Columbia V8P 1A1, Canada}

\author{Jillian M. Scudder}
\affiliation{Department of Physics and Astronomy, Oberlin College, Oberlin, Ohio, OH 44074, USA}

\author{Jorge Barrera-Ballesteros}
\affiliation{Department of Physics and Astronomy, Johns Hopkins University, Bloomberg Center, 3400 N. Charles St., Baltimore, MD 21218, USA}

\author{Fangting Yuan}
\affiliation{Shanghai Astronomical Observatory, Nandan Road 80, Shanghai 200030, China}

\author{Am\'elie Saintonge}
\affiliation{University College London, Gower Street, London WC1E 6BT, UK}

\author{Christine D. Wilson}
\affiliation{Department of Physics and Astronomy, McMaster University, Hamilton, ON L8S 4M1, Canada}

\author{Ho Seong Hwang}
\affiliation{Quantum Universe Center, Korea Institute for Advanced Study, 85 Hoegiro, Dongdaemun-gu, Seoul 02455, Korea}

\author{Ilse De Looze}
\affiliation{University College London, Gower Street, London WC1E 6BT, UK}
\affiliation{Sterrenkundig Observatorium, Ghent University, Krijgslaan 281 - S9, 9000 Gent, Belgium}

\author{Yu Gao}
\affiliation{Purple Mountain Observatory \& Key Laboratory for Radio Astronomy, Chinese Academy of Sciences, 8 Yuanhua Road, Nanjing 210034, China}

\author{Luis C. Ho}
\affiliation{Kavli Institute for Astronomy and Astrophysics, Peking University, Beijing 100871, China}
\affiliation{Department of Astronomy, School of Physics, Peking University, Beijing 100871, China}

\author{Elias Brinks}
\affiliation{Centre for Astrophysics Research, University of Hertfordshire, College Lane, Hatfield AL10 9AB, UK}

\author{Angus Mok}
\affiliation{Department of Physics \& Astronomy, The University of Toledo, Toledo, OH 43606, USA}

\author{Toby Brown}
\affiliation{Department of Physics and Astronomy, McMaster University, Hamilton, ON L8S 4M1, Canada}

\author{Timothy A. Davis}
\affiliation{School of Physics and Astronomy, Cardiff University, Queen's Building, The Parade, Cardiff CF24 3AA, UK}

\author{Thomas G. Williams}
\affiliation{School of Physics and Astronomy, Cardiff University, Queen's Building, The Parade, Cardiff CF24 3AA, UK}

\author{Aeree Chung}
\affiliation{Department of Astronomy, Yonsei University, 50 Yonsei-ro, Seodaemun-gu, Seoul 03722, Republic of Korea}

\author{Harriet Parsons}
\affiliation{East Asian Observatory, 660 N. A'ohoku Place, Hilo, HI 96720, USA}

\author{Martin Bureau}
\affiliation{Sub-Department of Astrophysics, University of Oxford, Denys Wilkinson Building, Keble Road, Oxford OX1 3RH, UK}

\author{Mark T. Sargent}
\affiliation{Astronomy Centre, Department of Physics and Astronomy, University of Sussex, Brighton BN1 9QH, UK}


\author{Eun Jung Chung}
\affiliation{Korea Astronomy and Space Science Institute, 776 Daedeokdaero, Yuseong-gu, Daejeon 34055, Republic of Korea}

\author{Eunbin Kim}
\affiliation{School of Space Research, Kyung Hee University, Yongin, Gyeonggi 17104, Republic of Korea}
\affiliation{Korea Astronomy and Space Science Institute, 776 Daedeokdaero, Yuseong-gu, Daejeon 34055, Republic of Korea}

\author{Tie Liu}
\affiliation{Korea Astronomy and Space Science Institute, 776 Daedeokdaero, Yuseong-gu, Daejeon 34055, Republic of Korea}
\affiliation{East Asian Observatory, 660 N. A'ohoku Place, Hilo, HI 96720, USA}
\affiliation{National Astronomical Observatories, Chinese Academy of Sciences, A20 Datun Road, Chaoyang District, Beijing 100012, China}

\author{Micha\l{} J. Micha\l{}owski}
\affiliation{Astronomical Observatory Institute, Faculty of Physics, Adam Mickiewicz University, ul. S\l{}oneczna 36, 60-286 Pozna\'{o}, Poland}

\author{Tomoka Tosaki}
\affiliation{Joetsu University of Education, Yamayashiki-machi, Joetsu, Niigata 943-8512, Japan}



\begin{abstract} 
Galaxy interactions are  often  accompanied by an  enhanced  star formation rate (SFR).
Since molecular gas is essential for star formation, it is vital to establish whether, and by how much, galaxy interactions affect the molecular gas properties.
We investigate the effect of  interactions on global molecular gas properties by studying a sample of 58 galaxies in pairs and 154 control galaxies.
Molecular gas properties are determined from observations with the  JCMT, PMO, CSO telescopes, and supplemented with data from the xCOLD GASS and JINGLE surveys at $^{12}$CO(1--0) and $^{12}$CO(2--1).
The SFR,  gas mass ($M_\mathrm{H_{2}}$),  and gas fraction  ($f_{gas}$) are all enhanced  in galaxies in pairs by $\sim$ 2.5 times    compared to the controls matched in redshift, mass, and effective radius, while the enhancement of star formation efficiency   (SFE $\equiv$ SFR/$M_{H_{2}}$) is less than a factor of 2.
We also find that the enhancements in SFR, $M_{H_{2}}$ and $f_{gas}$ increase with decreasing pair separation  and are larger in systems with smaller stellar mass ratio.
Conversely, the SFE is only enhanced in close pairs  (separation $<$ 20 kpc) and equal-mass systems; therefore most galaxies in pairs lie in the same parameter space on the  SFR-$M_{H_{2}}$ plane as controls.
This is the first time that the dependence of molecular gas properties on merger configurations is probed statistically with a relatively large sample and with a carefully-selected control sample for individual galaxies.
We conclude that galaxy interactions do modify the  molecular gas properties,  although the strength of the effect is merger configuration dependent.

\end{abstract} 


\section{Introduction}
It has been well established that galaxy interactions can trigger bursts of star formation.
Interaction-triggered star formation was first observed by \cite{Lar78}, who found that interacting galaxies have large dispersion in U-B/B-V colors due to the short duration starbursts.
Since then, many observations  have confirmed this finding \citep[e.g.,][]{Dar10a,Scu12,Kna15}.
Observationally,  the strongest starbursts (e.g., ultraluminous infrared galaxies, ULIRGs) are predominantly merging systems \citep{San88,Bor99}, which supports the idea that galaxy interactions are efficient in converting gas to stars.
Simulations also show  that  external perturbations can trigger star formation by the gas inflow induced as a result of  tidal forces \citep[e.g.,][]{Bar96,Dim07,Mor15}.

However, an enhanced star formation rate (SFR) is not  ubiquitous in interacting galaxies.
The average level of the SFR enhancement  of galaxy pairs, as measured in both observations and simulations, is moderate,  typically  below a factor of a few \citep{Dim07,Dim08,Mar08,Lin07,Hwa11,Won11,Xu12,Scu12,Ell13,Kna15}.
Star formation enhancement depends on parameters such as separation between galaxies in  pairs \citep{Lam03,Eli08,Par09,Hwa10,Scu12,Pat13,Dav15}, the properties of the progenitor galaxies \citep[][]{Mih96,Cox08,Xu10,Scu12,Dav15},  merging geometry \citep{Spr05,Dim07,Mor15,Spa16}, and gas (\ion{H}{1}) fraction \citep{Scu15}.

Given that SFR  depends on the molecular gas reservoir \citep{Ken98a}, one would expect that the amount or the physical properties of the molecular gas change while a galaxy undergoes an interaction  with another galaxy \citep[e.g.,][]{Mor18}.
Two possibilities for  enhanced star formation in galaxy pairs are most commonly proposed: (1) an enrichment of the molecular gas reservoir, which fuels star formation \citep[e.g.,][]{Com94,Cas04}  or (2)   an increase in the efficiency of converting gas into stars \citep[e.g.,][]{Sol88,Sof93,Mic16}. 
Both scenarios have observationally testable predictions --
the former  predicts higher molecular gas mass ($M_\mathrm{H_{2}}$), or, more precisely, higher molecular gas mass fraction with respect to the total (gas and stars) mass ($f_{gas}$), while  the latter  predicts higher star formation efficiency (SFE) of molecular gas.
  
However, observations of H$_{2}$ in interacting galaxies have  yet to give a clear picture of whether it is the total gas reservoir, or the SFE, that drives the enhanced SFR in galaxy pairs.
\cite{Sol88} observed $^{12}$CO(1-0) in  93 far-infrared bright pairs and classified them into five types according to the degree of interaction.
They found that there is no significant difference in CO luminosity (given that $L_\mathrm{CO}$ $\propto$ $M_\mathrm{H_{2}}$)  between their pair types and isolated galaxies.
On the other hand,  strong interactions give rise to an increase in  $L_\mathrm{FIR}$/$L_\mathrm{CO}$ ratio ($\propto$ SFE).
\cite{Sof93}  also found an  elevated SFE in their 54  interacting galaxies taken from the Arp Atlas.   
More recently, \cite{Mic16} revealed an increasing SFE from isolated to interacting galaxies and from early-stage to late-stage interactions using a sample of 60 interacting and 28 isolated galaxies\footnote{But note that they use $^{12}$CO(3-2) as molecular gas tracer, which may not trace total gas content due to the high critical density of $^{12}$CO(3-2) (a few times 10$^{3}$ cm$^{-3}$).}.
Yet several studies have arrived at the opposite conclusion. 
For example, \cite{Com94} observed $^{12}$CO(1-0) in 51 interacting galaxies and  find that  the total molecular content increases with decreasing projected separation of the pairs,  while SFE does not.
Accordingly, they concluded that the total molecular content plays a more significant role in triggering star formation than SFE.
A similar result is also reached by \cite{Cas04} using several hundreds of  interacting galaxies and $\sim$ 2000 normal galaxies compiled from the literature.
 
Although the above results indicate that galaxy interactions may affect the molecular gas properties, the above analyses have several shortcomings, which might contribute to their conflicting results.
For example,  global properties of the  control galaxies to be compared with pairs should be carefully controlled.
The majority of previous studies compare the properties of interacting and isolated galaxies directly, where the latter may not always be the perfect reference in terms of the distributions of their redshift, stellar mass  ($M_{\ast}$), and other galaxy properties. 
Another important factor is the choice of CO-to-H$_{2}$ conversion factor ($\alpha_\mathrm{CO}$) between the measured $L_\mathrm{CO}$ and $M_\mathrm{H_{2}}$. 
The validity of the widely-adopted Galactic $\alpha_\mathrm{CO}$ is often  questioned \citep[see the review by][]{Bol13}. 
Empirically, $\alpha_\mathrm{CO}$ increases with decreasing gas-phase metallicity 12$+$log(O/H) due to the decreasing abundance of CO relative to H$_{2}$ \citep{Ari96,Ler11,Nar12}.
Since gas-phase metallicity   is known to correlate with   $M_{\ast}$,  $M_\mathrm{H_{2}}$, and SFR of a galaxy \citep{Tre04,Man10,Bot16}, a physically-motivated $\alpha_\mathrm{CO}$  is essential for the study of molecular gas in galaxies.

Recently, \cite{Vio18} have taken a step towards addressing these improvements by considering  a control sample with properties matched to the galaxy pairs and by using a physically-motivated $\alpha_\mathrm{CO}$.
They found that  galaxy pairs have higher  both SFE and  $f_{gas}$  compared to the control sample.
However,  the investigation of the relation between merger configuration and gas properties is limited by the small sample size (11 galaxies in pairs) in \cite{Vio18}.
It remains untested about how the change in gas properties correlates with  details of galaxy interaction properties, such as pair separation and mass ratio.

In this paper, we study  molecular gas properties, which are calculated using a  physically-motivated value of $\alpha_\mathrm{CO}$, towards a sample of 58 galaxies in pairs. 
We  compare their star formation and molecular gas properties with a sample of carefully matched control galaxies. 
The sample uniquely covers  major and minor mergers (from equal mass merger to a ratio of $\sim$ 100),  widely-separated pairs and close pairs, primary (higher $M_{\ast}$) and secondary (lower $M_{\ast}$) galaxies in a pair.
This is the first time that the dependence of molecular gas properties on merger properties is probed statistically with a relatively large  sample and with a carefully-selected control sample for individual galaxies in pairs.

This paper is organized as follows. 
In Section \ref{sec_data}, we describe our pair identification and data used in the analysis. In Section \ref{result_absolute_prop} and \ref{sec_offset}, we first compare the SFR and molecular gas properties of the galaxies in pairs and control sample by comparing the medians of the two samples.
Next, in Section \ref{sec_rp} and \ref{sec_mr}, we   explore the dependence of SFR and gas properties on merger properties, including projected separation and the stellar mass ratio of the two galaxies in a pair.
In Section \ref{sec_discuss}, we discuss the potential driver of star formation in galaxies in pairs and the locus of our galaxies  in the   SFR-$M_\mathrm{H_{2}}$ relation.
The main results are summarized in Section \ref{sec_summary}.

Throughout this paper, we assume $\Omega_\mathrm{m}$ $=$ 0.3, $\Omega_\mathrm{\Lambda}$ $=$ 0.7, $H_{0}$ $=$ 70 km s$^{-1}$ Mpc$^{-1}$, and a Kroupa initial mass function (IMF) of stars \citep{Kro01}.

\section{Data}
\label{sec_data}
\subsection{Molecular Gas Observations of Galaxies in Pairs}
\label{sec_mo_pairs}
The pair sample in this work is  obtained either by our group or compiled from several surveys, summarized in Table \ref{tab_obs}.
The final sample  consists of  58 galaxies in pairs and 154 isolated galaxies  from which the control galaxies are drawn.
The galaxies in pairs we refer to here are galaxies with a  spectroscopic or morphological companion. 
In most of the cases discussed in this paper, the molecular gas observations were made toward one of the galaxies in a pair,  except a few close pairs.
Emission from the companion might be detected for the pairs with the smallest projected separations. 
The potential effect of this contamination will be discussed in Section \ref{sec_real}.
Details of the  sample selection, observations, and data reduction  are described  in this section.

\begin{table*}[]
	\begin{center}
		\caption{Summary of the observations.}
		\label{tab_obs}
		\begin{tabular}{cccccccc}
			\hline
			\hline
			& \multicolumn{1}{l}{} & \multicolumn{4}{c}{galaxies in pairs}                                                             & \multicolumn{1}{l}{} & pool of controls \\ \cline{1-1} \cline{3-6} \cline{8-8} 
			project    &                      & PI programs    & JINGLE         & JINGLE Pilot                       & xCOLD GASS     &                      & xCOLD GASS       \\
			number     &                      & 21             & 5              & 2+2+1                              & 27             &                      & 154              \\
			telescope &                      & JCMT           & JCMT           & JCMT/PMO/CSO                       & IRAM           &                      & IRAM             \\
			tracer &                      & $^{12}$CO(2-1) & $^{12}$CO(2-1) & $^{12}$CO(2-1)/(1-0)/(2-1)         & $^{12}$CO(1-0) &                      & $^{12}$CO(1-0)   \\
			beam size   & \multicolumn{1}{l}{} & 22$\arcsec$    & 22$\arcsec$    & 22$\arcsec$/52$\arcsec$/30$\arcsec$ & 22$\arcsec$    &                      & 22$\arcsec$ \\
			\hline    
		\end{tabular}
	\end{center}
\end{table*}

\subsubsection{Pair sample: JCMT observations (PI programs)}
\label{sec_sample_pi}
The molecular gas observations of about half of the galaxies in pairs were obtained through  our two PI programs on  the James Clerk Maxwell Telescope (JCMT) (project codes: M17AP060 and M17BP053; PI: H.-A. Pan).

The pair sample was selected from the  2779  galaxies in the fifth Product Launch (MPL-5, corresponding to SDSS DR13) of Mapping Nearby Galaxies at APO (MaNGA). 
MaNGA is part of the fourth generation of the Sloan Digital Sky Survey \citep[SDSS-IV; ][]{Gun06,Bla17} and aims to obtain  spatially resolved spectroscopy of 10,000 galaxies with median redshift $\sim$ 0.03 by 2020. MaNGA has a wavelength coverage of 3600 -- 10300\AA, with a spectral resolution varying from R $\sim$ 1400 at 4000 \AA\, to R$\sim$ 2600 around 9000 \AA. Further details on the science goals as well as sample selection can be found in \cite{Bun15} and \cite{Wak17}. 
While this work focuses on the  \emph{ globally integrated} star formation and molecular gas properties, the existence of  MaNGA data   will be beneficial in advancing the analysis of the spatially-resolved properties in the future.

We first identify galaxies in pairs in these  2779 MaNGA galaxies.
The galaxies in pairs are defined as those systems with projected separation ($r_{p}$) $<$ 50 kpc h$^{-1}$ (around 71.4 kpc with h $=$ 0.7) and line-of-sight velocity difference ($\Delta V$) $<$ 500 km s$^{-1}$ \citep[e.g.,][]{Pat02,Lin04}. 
It has been noticed that the SFR enhancement can extend out to 150 kpc \citep{Pat13}, although the enhancement for $r_{p}$ $>$ 100 kpc is almost negligible. 
Moreover, even with spectroscopic redshifts,  interlopers may still exist and become a more significant effect with larger separation. 
For these reasons, the criterion of 50 kpc h$^{-1}$  seems to be a reasonable choice.
662 galaxies in pairs were identified by these criteria.
However, if the two merger components are too close (normally late-stage mergers) to be deblended by SDSS or do not have two separate spectroscopic redshift measurements, they will not be identified as galaxies in pairs. 
To pick up those late-stage systems, we use the  ``P-merger'' parameter (weighted-merger-vote fraction) from  Galaxy Zoo  \citep{Dar10a,Dar10b}.
P-merger quantifies the probability  that an object is a merger, based on  visual inspection of large numbers of objects by  human classifiers.
P-merger ranges from 0, an object looks nothing like a merger, to 1, an object  looks  unmistakably so.
The criteria of P-merger $>$ 0.4 suggested by \cite{Dar10a,Dar10b} is applied to select galaxies in pairs.
The number of galaxies in pairs in the MaNGA sample increases to 736 by adding the Galaxy Zoo criterion.

The required observing time for JCMT $^{12}$CO(2-1) observations for each of the identified galaxies in pairs is estimated in the following way.
We  first calculate  the expected $M_\mathrm{H_{2}}$ from the Wide-field Infrared Survey Explorer (WISE) 12 $\mu$m luminosity ($L_{\mathrm{12\mu m}}$) using the  $L_{\mathrm{12\mu m}}$-$M_\mathrm{H_{2}}$ relation  proposed by  \cite{Jia15}.
Since 12 $\mu$m  emission  is a good tracer of star formation \citep{Don12,Lee13},  the  relation is essentially the same  as the  Kennicutt-Schmidt relation.
The $^{12}$CO(1-0) luminosity  is  calculated from $M_\mathrm{H_{2}}$ using  the Milky Way value for $\alpha_{\mathrm{CO}}$ of 4.3  M$_{\odot}$ (K km s$^{-1}$ pc$^{2}$)$^{-1}$ \citep{Bol13}.
Subsequently,   the luminosity and flux of $^{12}$CO(2-1) is derived assuming a $^{12}$CO(2-1) to $^{12}$CO(1-0) line intensity ratio ($R_{21}$) of 0.6.
The adopted  $R_{21}$ here is a conservative choice (lower limit) for time estimation, as $R_{21}$ is found to be 0.6 -- 1.0 in nearby galaxies \citep[e.g.,][]{Bra92,Ler09}.
Although we made   conservative assumptions for $R_{21}$ and $\alpha_{\mathrm{CO}}$ for the purpose of estimating the exposure time, later we will use a different $R_{21}$ which is chosen based on  gas properties; moreover, we will  present a physically-motivated $\alpha_{\mathrm{CO}}$ prescription that computes the value on a galaxy by galaxy basis in Section \ref{result_absolute_prop}, and will  explore these assumptions in Section \ref{sec_caveats}.
Finally, we  estimate the required time for a 4$\sigma$ detection with  velocity resolution of 30 km s$^{-1}$   for each identified galaxies in pairs and propose to observe 41 galaxies that have suitable  Declination range and require on-source time $<$ 250 minutes each.
Any possible bias introduced by this latter choice will be discussed in Section \ref{sec_real}.

$^{12}$CO(2--1) (230.538 GHz) observations of the 41 galaxies in pairs  were obtained by JCMT  using the RxA3m receiver (but only 21 galaxies are used for the analysis after a further control on galaxy properties, we will show the criteria in the next paragraph). 
The redshift range of these galaxies in pairs is 0.02 -- 0.06.
The beam size of the telescope is 22$\arcsec$ at 230 GHz. 
The observations were conducted during several periods from   November 2016 to January 2018.
The sky opacity at 225GHz was reported by the JCMT's Water Vapour Monitor (WVM), an in-cabin line-of-sight radiometer assessing the 183 GHz water line (reported at 225GHz for historic reasons).
The typical sky opacity was 0.08 -- 0.20.
The typical system temperatures  were between 200 and 400 K. 
The on-source time for individual galaxies ranged from 40 minutes to 4 hours. 
The total on-source time for the two PI programs  was $\sim$ 90 hours.
The data reduction was done using the Starlink software \citep{Cur14}.
Individual exposures  ($\sim$ 20 -- 40 min., including calibration)  of a given galaxy were calibrated separately, and then coadded.
The spectrum was binned to a velocity resolution of 30 km s$^{-1}$. 
A linear baseline was subtracted from the  spectrum using line-free channels.  
For a few galaxies  for which the  baseline is structured,  a second- or third-order polynomial was used to subtract the baseline. 
Spectra were converted from antenna temperature in K to Jy by applying a factor of $15.6/\eta_a$, where the aperture efficiency $\eta_a$ is 0.55. 
The integrated CO luminosity $L_{\mathrm{CO}}$ is computed   following \cite{Sol97},
\begin{equation}
\frac{L_{\mathrm{CO}}}{\mathrm{[K\;km s^{-1}\; pc^{2}]}}=3.27\times 10^{7}S\mathrm{_{CO}}\nu _{\mathrm{CO}}^{-2}D_{L}^{2}(1+z)^{-3}.
\end{equation}
In this expression,  $S\mathrm{_{CO}}$ is the line flux in unit of Jy km s$^{-1}$, $\nu _{\mathrm{CO}}$ is the observed frequency in GHz, $D_{L}$ represents the luminosity distance  in Mpc.
For $^{12}$CO (2--1) observations, a  $^{12}$CO (2-1)-to-$^{12}$CO(1-0)  intensity ratio of  0.8 is assumed when calculating $L_\mathrm{CO}$ \citep{Ler09}.
If  the gas is optically thick, a ratio of 0.8  corresponds to an excitation temperature of $\sim$ 10 K.
A detection (i.e., signal-to-noise ratio $>$ 3) rate of $\sim$ 90\% is achieved, implying that the approach we have taken to calculate the required sensitivity is valid.  

Although 41 galaxies in pairs were obtained from the  PI programs, only 21 galaxies are used for the analysis in this paper.
Since the control sample used in this work (\S\ref{sec_control_sample}) has a stellar mass cut of log($M_{\ast}$/M$_{\odot}$) $=$ 9 \citep{Sai17}, we remove galaxies that have $M_{\ast}$ less than this value.
Moreover, only spiral galaxies are used, because early-type galaxies potentially have lower SFR, $M_\mathrm{H_{2}}$, and $f_{gas}$, which may be irrelevant to the existence of interaction or not.
The Galaxy Zoo property, debiased probability of being a spiral galaxy ``P-CS'', is used to identify galaxy morphology. 
We select  galaxies that have P-CS $\geq$ 0.6.
All of the 21 galaxies in pairs have a solid detection by JCMT.
Of these, 20 are selected based on pair separation and one is based on the Galaxy Zoo morphology.

\subsubsection{Pair sample: JINGLE}
\label{sec_sample_jingle}
The JCMT dust and gas In Nearby Galaxies Legacy Exploration (JINGLE) is an ongoing JCMT Large Program \citep{Sai18}. 
JINGLE is designed to systematically study the cold interstellar medium of galaxies in the local Universe.
The survey observed 850 $\mu$m dust continuum  with SCUBA-2 for a  sample of 193 Herschel-selected galaxies with  log($M_{\ast}$/M$_{\odot}$) $>$ 9, and integrated $^{12}$CO(2-1) line fluxes with RxA3m for a subset of 97 of these galaxies.
63 out of the 97 galaxies are  within the footprint of MaNGA.

We briefly summarize the sample selection of JINGLE here.
The JINGLE parent sample consists of  $\sim$ 2800 galaxies with log($M_{\ast}$/M$_{\odot}$) $>$ 9 and 0.01 $<$ $z$ $<$ 0.05 within  the North Galactic Pole (NGP) region and three of the equatorial Galaxy And Mass Assembly (GAMA) fields (GAMA09, GAMA12 and GAMA15).
The sample is narrowed down to $\sim$ 280 galaxies with   a $>$ 3$\sigma$ detection at   both 250 and 350$\mu$m in the Herschel ATLAS survey,  and  are predicted to be detectable with SCUBA-2 in less than 2 hours of integration.
Then  193 galaxies are selected in order to have a uniform stellar mass distribution at log($M_{\ast}$/M$_{\sun}$) $>$ 9.
A sub-sample of 97 galaxies  predicted to be detectable in on-source time of 345 minutes are selected to obtain integrated $^{12}$CO(2-1) line fluxes.
Two methods are used to estimate the flux and  integration time for  $^{12}$CO(2-1) observations.
The first method is the same as that used for the PI programs in Section \ref{sec_sample_pi}.
The second approach  is based on the 2 Star Formation Mode  formal flux prediction of \citet{Sar14}, in which the $^{12}$CO(1-0) line flux is related to the  galaxy's position in SFR-$M_{\ast}$ plane.
The predicted fluxes from the two methods agree well with each other.
We note that JINGLE adopts an $R_\mathrm{21}$ of 0.7 for the required sensitivity and observing time estimation.
More details on the JINGLE design, as well as the sample selection and science goals are given by \cite{Sai18} (overview of the survey), Smith et al. in prep. (details of dust observations), and Xiao et al. in prep. (details of molecular gas observations).

We apply the same criteria  to select galaxies in pairs as described in Section \ref{sec_sample_pi} ($r_{p}$ $<$ 50 kpc h$^{-1}$ and $\Delta V$ $<$ 500 km s$^{-1}$, or P-merger $>$ 0.4,  and P-CS $>$ 0.6) to the 45 JINGLE galaxies for which CO data has been obtained before August 2017\footnote{For reference, the number of galaxies in first CO data release will be 72  (Xiao et al. in prep.).}.
A total of 5  galaxies in pairs are identified in this way, all of them  are identified through the $r_{p}$ and $\Delta V$ criteria. 
The data reduction is carried out in the same way as our PI programs described in Section \ref{sec_sample_pi}.

\subsubsection{Pair sample: JINGLE Pilot}
\label{sec_sample_jinglep}
The JINGLE Pilot program (Gao et al.  in prep.) is a series of $^{12}$CO (2-1) and (1-0) observations of MaNGA galaxies carried out by multiple facilities including JCMT, the 14-m telescope of the Purple Mountain Observation (PMO), and the 10.4-m telescope of the Caltech Submillimeter Observatory (CSO).
The project was designed as a test of the JINGLE survey.

Galaxies were selected from the MaNGA MPL-3   (720  galaxies).
The CO flux estimation for the MPL-3 galaxies also made use of the $L_{\mathrm{12\mu m}}$-$M_\mathrm{H_{2}}$ relation as described in Section \ref{sec_sample_pi}.
A sample of 31 galaxies were selected for observations.
The redshift range of these galaxies is 0.02 -- 0.04. 
The galaxies were  assigned to the various telescopes listed above, according to the required sensitivity and the  sensitivities of the telescopes.
Some galaxies were observed by multiple telescopes to obtain both  $^{12}$CO(1-0) and $^{12}$CO (2-1) data. 
The multiple transitions can be used to trace the physical conditions (e.g., temperature and density) of molecular gas.

There were 21 galaxies assigned to be observed by JCMT at  $^{12}$CO(2-1).
The observations were done with the RxA receiver\footnote{Observations taken at 230 GHz at the JCMT prior to December 2015 were taken with RxA. Observations taken after this date (specifically, after May 13th of 2016) are observed with a replacement mixer. The JCMT thus calls the new instrument RxA3m.} between  March to  November of 2015, with a typical sky opacity of $\sim$0.12 to 0.32.
A total of 17 galaxies  were assigned to the PMO 14-m telescope at $^{12}$CO(1-0). 
Observations were carried out in the winter of 2015 with the nine-beam receiver.
The beam size  of PMO observations at 115 GHz is 52$\arcsec$, which can cover the entire  galaxy at the typical redshift of the sample.
Three galaxies were observed by CSO at $^{12}$CO(2-1) with  the Heterodyne receiver in February of 2015.
The beam size of CSO at 230 GHz is 30$\arcsec$.
The full design of the project and details of the data reduction are presented in Gao et al.  in prep..

From a total of 31 galaxies in the JINGLE Pilot sample,  we identify 5 additional galaxies in pairs for our analysis   based on the criteria described in  Section \ref{sec_sample_pi}, with 2 from JCMT, 2 from PMO and 1 from CSO observations.
Of these, 4 are selected based on $r_{p}$ and $\Delta V$ and 1 based on Galaxy Zoo classification.

\subsubsection{Pair sample: xCOLD GASS}
\label{sec_sample_xcg}
We also include galaxies in pairs from  The Extended CO Legacy Database for GASS  \citep[xCOLD GASS;][]{Sai17}.
xCOLD GASS is an extension of the IRAM 30-m legacy survey COLD GASS \citep{Sai11}, which studies the molecular gas of nearby late-type galaxies with stellar masses 10 $<$ log($M_{\ast}$/M$_{\odot}$) $<$ 11.5 and redshift 0.0025 $<$ $z$ $<$ 0.05.
xCOLD GASS  extends the sample to  log($M_{\ast}$/M$_{\odot}$) $=$ 9.0.
COLD GASS and xCOLD GASS observe galaxies in $^{12}$CO(1-0) and $^{12}$CO(2-1) with IRAM and with complementary observations from APEX $^{12}$CO(2-1).
$^{12}$CO(1-0) data are used for this work because the beam size is exactly the same as for the JCMT 15-m telescope at 230 GHz.

We identify 27 galaxies in pairs from all 532 xCOLD GASS galaxies.
Criteria for selecting galaxies in pairs are the same as described in Section \ref{sec_sample_pi} ($r_{p}$ $<$ 50 kpc h$^{-1}$ and $\Delta V$ $<$ 500 km s$^{-1}$, or P-merger $>$ 0.4, and P-CS $>$ 0.6).
A total of 27 galaxies in pairs are identified, and all of them  are identified through the $r_{p}$ and $\Delta V$ criteria. 

We have checked whether the galaxies from various observations reach different depth in terms of sensitivity.
The main difference between the sample selection for the PI program/JINGLE/JINGLE Pilot and  xCOLD GASS is the sensitivity (integration time) estimation.
The former ones use the value of SFR/$M_\mathrm{H_{2}}$, while the latter  uses  $M_\mathrm{H_{2}}$/$M_{\ast}$.
The $M_\mathrm{H_{2}}$/$M_{\ast}$ limit  is 2.5 per cent for xCOLD GASS \citep{Sai17}.
For the galaxies from the PI programs, JINGLE and JINGLE Pilot programs, the typical $M_\mathrm{H_{2}}$/$M_{\ast}$ achieved is 3.3 per cent,  assuming $R_{21}$ = 0.6, $\alpha_\mathrm{CO}$ = 4.3 (as the values used in \S\ref{sec_sample_pi}), and a common line width of 300 km s$^{-1}$.
We thus conclude that the samples analysed here have comparable depth.

\subsubsection{Pair sample: Summary}

In summary, our  sample consists of a total of 58 galaxies in pairs (Table \ref{tab_gal}), of which 21 are from the JCMT PI programs, 5 from JINGLE,  5 from the JINGLE Pilot program, and 27 from xCOLD GASS.
56 galaxies are selected based on $r_{p}$ and $\Delta V$ and   2 (1 from the PI program and 1 from the JINGLE Pilot program) based on Galaxy Zoo morphologies.
It is important to note that the Galaxy Zoo classification could potentially pick up galaxies in the post-coalescence  stage (i.e., post-merger).
This is not the case for the two galaxies identified through their morphologies; in other words,  our sample does not contain post-mergers.
We refer the reader to \cite{Ell13}, \cite{Ell18}, \cite{Tho18}, and Sargent et al. in prep. for the star formation and cold gas properties of post-mergers.

\begin{figure}%
	\centering
	\includegraphics[width=0.46\textwidth]{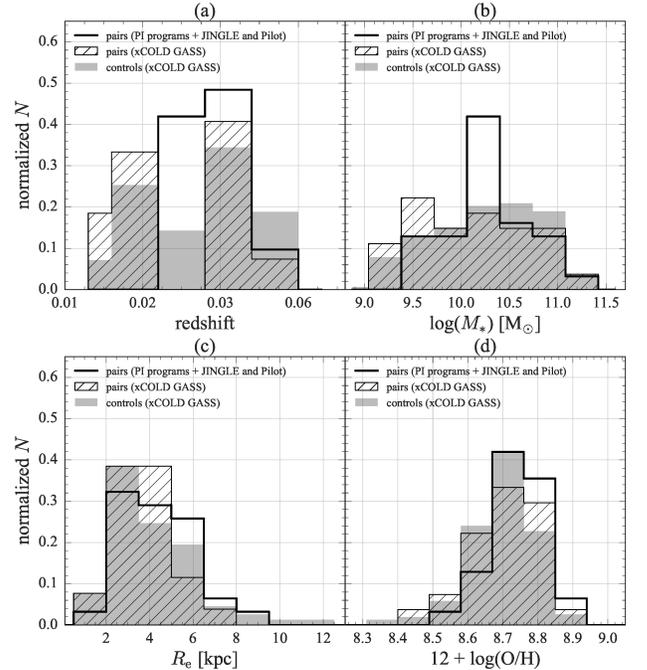}
	\caption{Histograms showing the distribution of the sample galaxies in terms of redshift (a), stellar mass (b), effective radius measured at $r$-band (c), and gas-phase metallicity (d). The open histograms represent galaxies in pairs from the PI programs, JINGLE, and JINGLE Pilot. The hatched histograms show the galaxies in pairs from the xCOLD GASS survey. The pool of controls are shown as filled histograms.  }%
	\label{fig_gal_props}%
\end{figure}

\begin{figure*}%
	\centering
	\includegraphics[angle=-90,width=0.8\textwidth]{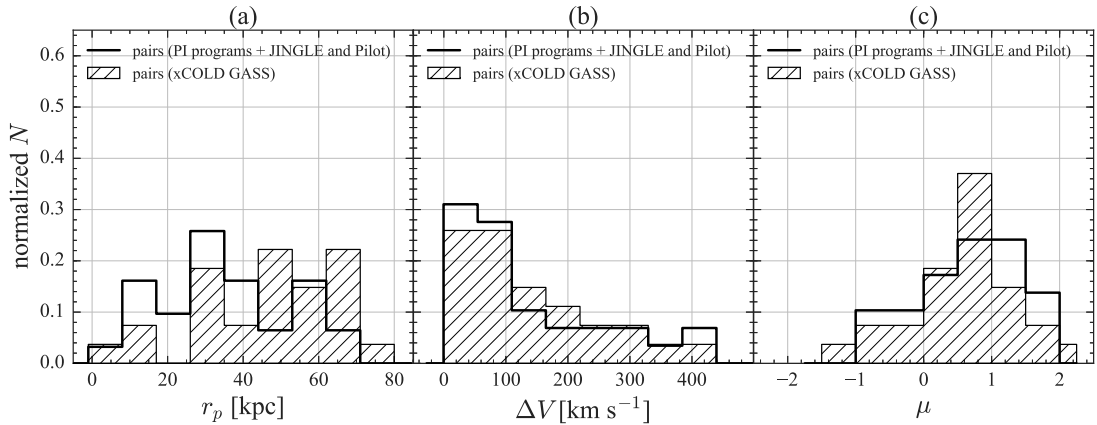}
	\caption{Distributions of merger configurations across the galaxies in pairs from the PI programs, JINGLE, and JINGLE Pilot (open histograms) and xCOLD GASS (hatched histograms): (a) projected separation  of the two galaxies, (b) line-of-sight velocity difference, and (c) stellar mass ratio.} %
	\label{fig_merger_props}%
\end{figure*}

For the two galaxies identified through Galaxy Zoo morphologies,  we  estimate their $r_{p}$ by calculating the distance between the two galactic nuclei. 
The distributions of galaxy properties are shown in Figure \ref{fig_gal_props}.
Their merger properties are presented in Figure \ref{fig_merger_props}  (open and hatched histograms).

\LTcapwidth=\textwidth
\begin{center}
	
	\begin{longtable*}{ccccccccccc}
	\caption{
		Table lists the  physical properties of galaxies in pairs: (1) SDSS ID, (2) SDSS spectroscopic redshift, (3) stellar mass from the MPA-JHU catalog (Section \ref{sec_sfr_sm}), (4) SFR from  the MPA-JHU catalog (Section \ref{sec_sfr_sm}), (5) NSA 50\% light radius measured at $r$-band, (6) projected separation between two galaxies in a pair, (7) stellar mass ratio between two galaxies in a pair (Section \ref{sec_mr_cal}), (8) aperture-corrected and line-ratio-corrected ($R_{21}$ $=$ 0.8) CO luminosity $L_\mathrm{CO}$ and its uncertainty (Section \ref{sec_ap_cor}), (9) gas-phase metallicity (Section \ref{sec_oh}), (10) CO-to-H$_{2}$ conversion factor (Section \ref{result_absolute_prop}), and (11) parent samples (Section \ref{sec_mo_pairs}), where J $=$ JINGLE (\S\ref{sec_sample_jingle}), xCG $=$ xCOLD GASS (\S\ref{sec_sample_xcg}), PI $=$ PI programs (\S\ref{sec_sample_pi}), and JP $=$ JINGLE Pilot (\S\ref{sec_sample_jinglep}).
	} \\ \toprule
	\label{tab_gal} 

Source & $z$ & log($M_{\ast}$) & log(SFR) & $R_\mathrm{e}$ & $r_{p}$ & $\mu$ & $L_\mathrm{CO}$/10$^{8}$ & 12+log(O/H)  & $\alpha_\mathrm{CO}$ & parent\\
 &   & [M$_{\odot}$] & [M$_{\odot}$ yr$^{-1}$]& [kpc] & [kpc] &  &  [K km s$^{-1}$ pc$^{2}$] & & [M$_{\odot}$/$L_\mathrm{CO}$] & sample \\
 (1) &(2) &(3) &(4)&(5)&(6)&(7)&(8)&(9)&(10)& (11)\\
		\hline
	 \endfirsthead
	 
	 \toprule

Source & $z$ & log($M_{\ast}$) & log(SFR) & $R_\mathrm{e}$ & $r_{p}$ & $\mu$ & $L_\mathrm{CO}$/10$^{8}$ & 12+log(O/H)  & $\alpha_\mathrm{CO}$ & parent\\
&   & [M$_{\odot}$] & [M$_{\odot}$ yr$^{-1}$]& [kpc] & [kpc] &  &  [K km s$^{-1}$ pc$^{2}$] & & [M$_{\odot}$/$L_\mathrm{CO}$] & sample \\
(1) &(2) &(3) &(4)&(5)&(6)&(7)&(8)&(9)&(10)& (11)\\ 
\hline 
\endhead 

J130125.07+284038.0 & 0.029 & 10.23 & 0.33 & 8.3 & 27.88 & 1.76 & 7.39(0.49) & 8.82 & 2.22 & J \\
J130615.12+252737.9 & 0.024 & 10.12 & 0.37 & 6.1 & 16.81 & 0.63 & 7.71(0.91) & 8.76 & 2.85 & J \\
J132035.40+340821.7 & 0.023 & 10.29 & 0.82 & 8.6 & 31.61 & 1.28 & 89.94(1.0) & 8.71 & 3.62 & J \\
J132443.68+323225.0 & 0.04 & 10.84 & 1.07 & 9.0 & 60.18 & 0.88 & 29.42(2.97) & 8.75 & 3.07 & J \\
J133457.27+340238.7 & 0.024 & 10.56 & 0.64 & 12.4 & 40.94 & 0.19 & 30.91(2.23) & 8.86 & 1.91 & J \\
 &  &  &  &  &  &  &  &  &  &  \\[-7pt]
J075641.84+175928.2 & 0.041 & 10.57 & 1.03 & 4.9 & 26.61 & 0.87 & 14.16(1.46) & 8.76 & 3.0 & xCG \\
J081115.92+251045.7 & 0.014 & 9.62 & -0.38 & 7.8 & 29.08 & 0.74 & 0.52(0.08) & 8.65 & 4.25 & xCG \\
J081905.10+214729.0 & 0.015 & 10.08 & 0.11 & 62.0 & 62.12 & 0.62 & 54.17(0.44) & 8.69 & 3.73 & xCG \\
J084256.38+133829.7 & 0.017 & 9.68 & 0.31 & 12.4 & 4.95 & 0.04 & 2.3(0.21) & 8.7 &3.83 & xCG \\
J085254.99+030908.4 & 0.035 & 10.28 & -0.04 & 4.0 & 63.95 & 1.1 & 1.58(0.32) & 8.61 & 5.02 & xCG \\
J090311.25+100907.0 & 0.03 & 10.11 & 0.3 & 9.6 & 51.6 & 0.53 & 5.83(0.58) & 8.84 &2.11 & xCG \\
J093236.58+095025.9 & 0.049 & 10.86 & 0.67 & 5.1 & 71.02 & 0.04 & 9.77(1.35) & 8.52 & 7.29 & xCG \\
J095324.56+074956.2 & 0.039 & 10.67 & -0.36 & 4.2 & 61.17 & 0.82 & 2.44(0.54) & 8.68 & 3.51 & xCG \\
J103333.43+115216.9 & 0.034 & 10.59 & 0.76 & 5.4 & 12.99 & 0.63 & 16.39(1.46) & 8.79 & 2.61 & xCG \\
J104024.66+065137.7 & 0.03 & 10.89 & 0.07 & 6.8 & 68.8 & 0.31 & 8.91(0.78) & 8.55 & 6.11 & xCG \\
J112746.27+265734.5 & 0.033 & 10.6 & -0.94 & 4.6 & 63.77 & 1.64 & 2.02(0.41) & 8.62 & 4.22 & xCG \\
J112946.35+152001.1 & 0.037 & 11.02 & -0.76 & 6.3 & 70.29 & 1.64 & 4.88(0.74) & 8.67 & 3.32 & xCG \\
J113116.03+043908.7 & 0.033 & 10.09 & 0.47 & 7.1 & 47.21 & 1.07 & 6.51(0.65) & 8.87 & 1.92 & xCG \\
J113701.89+153414.1 & 0.013 & 9.88 & -0.52 & 15.4 & 45.25 & 1.44 & 1.84(0.16) & 8.76 & 2.71 & xCG \\
J113914.72+145932.7 & 0.014 & 9.64 & -0.36 & 5.1 & 47.96 & 1.21 & 0.58(0.08) & 8.71 & 3.4 & xCG \\
J115020.17+255742.7 & 0.013 & 9.36 & -0.21 & 11.7 & 30.97 & 0.88 & 0.49(0.07) & 8.44 & 10.33 & xCG \\
J115726.68+251359.0 & 0.015 & 9.37 & -0.73 & 7.1 & 48.58 & -1.15 & 0.45(0.09) & 8.76 & 2.75 & xCG \\
J120222.51+295142.3 & 0.01 & 9.98 & -0.16 & 8.8 & 55.07 & 0.76 & 5.91(0.5) & 8.77 & 2.69 & xCG \\
J120409.73+014933.5 & 0.017 & 9.67 & -0.21 & 10.5 & 31.51 & -0.9 & 1.18(0.18) & 8.8 & 2.41 & xCG \\
J125905.29+273839.9 & 0.018 & 9.67 & 0.09 & 6.6 & 40.87 & -0.26 & 1.98(0.21) & 8.82 & 2.32 & xCG \\
J130750.80+031140.7 & 0.039 & 11.12 & -0.42 & 7.9 & 43.51 & 2.03 & 7.41(0.78) & 8.7 & 3.04 & xCG \\
J134701.23+335336.9 & 0.017 & 9.78 & -0.31 & 16.0 & 26.29 & 0.29 & 1.21(0.15) & 8.64 & 4.46 & xCG \\
J135655.41+140832.1 & 0.015 & 9.31 & -0.81 & 9.5 & 59.52 & 0.9 & 0.64(0.11) & 8.64& 4.28 & xCG \\
J142342.38+340032.4 & 0.013 & 9.84 & -0.02 & 8.4 & 14.95 & -0.18 & 2.46(0.2) & 8.77 & 2.72 & xCG \\
J143525.34+002003.5 & 0.035 & 10.2 & 0.71 & 3.7 & 66.43 & 0.89 & 12.2(1.11) & 8.79& 2.66 & xCG \\
J225258.55+010833.3 & 0.016 & 9.5 & -0.7 & 10.3 & 58.37 & -0.87 & 0.58(0.08) & 8.6& 5.12 & xCG \\
J231229.22+135632.1 & 0.034 & 10.91 & -0.49 & 5.4 & 51.67 & 0.37 & 8.23(0.92) & 8.67 & 3.51 & xCG \\

 &  &  &  &  &  &  &  &  &  &  \\[-7pt]

J025057.46+002209.8 & 0.044 & 10.05 & 0.63 & 3.1 & 26.08 & -0.13 & 3.34(0.79) & 8.8 & 2.58 & PI \\
J031943.04+003355.7 & 0.024 & 10.06 & 0.27 & 7.6 & 44.48 & -0.75 & 1.83(0.66) & 8.73 & 3.19 & PI \\
J032043.18-010008.2 & 0.036 & 10.64 & 0.68 & 5.5 & 12.15 & 1.14 & 35.82(1.69) & 8.75 & 2.97 & PI \\
J032247.22+000857.7 & 0.023 & 10.38 & -1.29 & 13.7 & 32.19 & 1.18 & 6.77(0.77) & 8.67 & 3.38 & PI \\
J075454.46+535046.5 & 0.035 & 10.76 & 0.7 & 6.0 & 60.07 & 1.69 & 21.41(1.21) & 8.72 & 3.4 & PI \\
J082150.16+453110.6 & 0.054 & 10.38 & 0.77 & 3.4 & 59.25 & 0.73 & 14.95(1.02) & 8.85 & 2.07 & PI \\
J093846.17+483346.3 & 0.025 & 9.43 & 0.11 & 5.4 & 20.88 & -0.94 & 3.1(0.43) & 8.71& 3.6 & PI \\
J100508.31+443050.5 & 0.026 & 10.36 & -0.74 & 9.6 & 65.66 & 0.38 & 3.65(1.37) & 8.66 & 3.75 & PI \\
J100718.98+463247.1 & 0.024 & 10.17 & 0.03 & 7.1 & 37.5 & 1.44 & 5.14(0.72) & 8.82& 2.17 & PI \\
J102843.06+395019.9 & 0.029 & 9.95 & 0.45 & 3.5 & 38.56 & 0.53 & 4.22(1.25) & 8.82& 2.37 & PI \\
J102855.10+395341.3 & 0.044 & 10.26 & -0.0 & 6.2 & 27.37 & 0.5 & 13.15(1.07) & 8.71 & 3.32 & PI \\
J121049.28+443045.3 & 0.023 & 10.18 & -0.12 & 5.3 & 41.8 & 1.5 & 1.16(0.14) & 8.73& 3.09 & PI \\
J130420.70+450323.9 & 0.028 & 9.66 & 0.34 & 7.2 & 62.3 & 0.63 & 5.07(0.67) & 8.65 & 4.6 & PI \\
J134109.43+231640.5 & 0.027 & 9.88 & -0.16 & 3.1 & 13.98 & -0.66 & 12.21(1.31) & 8.75 & 2.94 & PI \\
J135129.47+434823.1 & 0.033 & 10.16 & 0.61 & 4.4 & 43.84 & 1.33 & 6.77(1.0) & 8.71& 3.62 & PI \\
J140057.82+425120.3 & 0.032 & 10.67 & 0.49 & 6.0 & 25.33 & 0.0 & 13.17(1.04) & 8.76 & 2.84 & PI \\
J153545.82+445005.2 & 0.03 & 10.19 & 0.47 & 8.8 & 49.31 & 1.03 & 8.9(1.35) & 8.78 & 2.66 & PI \\
J154219.34+475636.7 & 0.037 & 9.65 & -0.36 & 3.7 & 15.87 & -0.33 & 12.28(0.79) & 8.67 & 3.94 & PI \\
J160242.58+411150.1 & 0.033 & 10.41 & 0.69 & 7.8 & 17.46 & 0.42 & 22.59(2.2) & 8.82 & 2.31 & PI \\
J163349.62+391547.5 & 0.032 & 10.74 & 0.42 & 6.2 & 32.83 & 1.01 & 10.68(0.8) & 8.73 & 3.1 & PI \\
J172823.84+573243.4 & 0.029 & 9.61 & 0.8 & 3.9 & 7.79 & $\dots$ & 5.48(0.81) & 8.51& 8.47 & PI \\

 &  &  &  &  &  &  &  &  &  &  \\[-7pt]
J032042.95-010631.1 & 0.021 & 10.02 & -0.04 & 11.6 & 55.22 & 0.9 & 3.92(0.26) & 8.79 & 2.47 & JP \\
J074637.71+444725.8 & 0.031 & 11.16 & 0.2 & 9.3 & 33.58 & 0.6 & 39.69(2.53) & 8.67& 3.8 & JP \\
J091500.75+420127.8 & 0.028 & 10.27 & 0.87 & 4.9 & 61.02 & -0.35 & 9.48(1.85) & 8.76 & 3.01 & JP \\
J091555.53+441957.9 & 0.04 & 10.95 & 1.23 & 10.4 & 9.73 & $\dots$ & 46.21(4.58) & 8.78 & 2.81 & JP \\
J110637.36+460219.5 & 0.025 & 10.44 & 0.52 & 4.3 & 27.03 & 1.76 & 14.81(1.65) & 8.77 & 2.76 & JP \\

		\hline
	\end{longtable*}
\end{center}

\subsection{Control sample: xCOLD GASS}
\label{sec_control_sample}
In order to quantify the effect of galaxy interactions on star formation and molecular gas properties, isolated galaxies are used as a comparison sample to our galaxies in pairs.
Isolated galaxies are selected from xCOLD GASS and are systems that  have no  spectroscopic companion within $r_{p}$ $<$ 50 kpc h$^{-1}$ and $\Delta V$ $<$ 500 km s$^{-1}$ and have a P-merger value equal to zero.
We use the same criteria to control the morphology of the control sample as described in \S\ref{sec_sample_pi}.
Combining these criteria yields a control sample pool of 154 galaxies.
About 8\% of the selected control galaxies have no detection in $^{12}$CO(1-0). 
An upper limit for the flux of 3$\sigma$ is given for these galaxies \citep{Sai17}.
Since the integration limit $M_\mathrm{H_{2}}$/$M_{\ast}$ is as deep as  2.5 per cent for xCOLD GASS,  these non-detected galaxies  thus truly have lower gas fraction compared to other galaxies.
These galaxies with upper limits for $L_\mathrm{CO}$ are included in the analysis.
All of our conclusions remain unchanged if we use only galaxies with detections.
The distributions of galaxy properties of controls are shown in Figure \ref{fig_gal_props} (filled histograms).

\subsection{Aperture Correction}
\label{sec_ap_cor}

Some of the galaxies have optical sizes in excess of the telescope beams, so an aperture correction is required to correct the CO fluxes measured and turn them into estimates for  the total flux.
For reference,  range in the effective radius ($R_\mathrm{e}$: Petrosian half-light radius measured at $r$-band) of our sample is $\sim$ 2 -- 6 kpc as shown in Figure \ref{fig_gal_props}(c), corresponding to about 4$\arcsec$ -- 12$\arcsec$ at the  redshifts of our sample (Figure \ref{fig_gal_props}(a)).

Aperture corrections are applied to all galaxies in pairs and control galaxies in this study.
We adopt the method of \cite{Sai12}.
For each galaxy, we create a model galaxy having an exponential molecular gas distribution  with a profile following that of the stellar light.
This assumption is based on the observation that CO and SFR distributions trace each other well in nearby  galaxies \citep{Ler09}.
Then the model is convolved with a Gaussian  matching the properties of the telescope beams.
The aperture correction is the ratio between the total flux of the  model and the flux in the beam area.
The median aperture correction to the CO luminosity across the galaxies in pairs and the pool of controls  are 0.09 and 0.08 dex, receptively.

\subsection{Global Stellar Mass and Star Formation Rate}
\label{sec_sfr_sm}
The global SFR and $M_{\ast}$ are taken from the MPA-JHU DR7 public catalog\footnote{https://wwwmpa.mpa-garching.mpg.de/SDSS/DR7/\#derived}.
The MPA-JHU catalog assumes a Kroupa IMF \citep{Kro01}.
$M_{\ast}$ is estimated by fitting stellar population models from \cite{Bru03} to the $ugriz$ SDSS  photometry, following the method of \cite{Kau03}.
The $M_{\ast}$   have been found to agree with other estimates \citep[e.g.,][]{Tay11,Men14,Cha15}.
To estimate SFRs, \cite{Bri04} first distinguish the emission line properties based on  the theoretical upper (lower) limit for pure starburst (AGN) models  \citep{Kew01,Kau03b} on the Baldwin-Phillips-Terlevich (BPT)    diagram \citep{Bal81}.
For galaxies   in which the primary source of ionizing photons is from HII regions, SFRs are estimated by fitting a grid of photo-ionization models from \cite{Cha01} to the observed H$\alpha$, H$\beta$, [\ion{O}{3}],  and [\ion{N}{2}] line fluxes.
These SFR estimates agree well with those derived from
the infrared fluxes \citep{Cha02}.
For galaxies  falling outside of the star-forming regime on the BPT diagrams, since the line fluxes might be affected by the AGN component, their SFRs are estimated based on the relation between specific SFR (sSFR $=$ SFR/$M_{\ast}$) and D4000 \citep{Bri04}.
This relationship was constructed using the sSFR and D4000 of  star-forming galaxies.
According to the BPT diagram, the pair sample consists of 8 AGNs, 11 Composite, and 39 star-forming galaxies.
The pool of the control sample includes 31 AGNs, 34 Composite, and 89 star-forming galaxies.
It is worth noting that by adopting currently popular modified-BPT diagrams to distinguish Seyfert and low-ionization emission-line (nuclear) regions (LI(N)ERs) in the AGN regime \citep{Cid13}, the majority of the AGNs in our sample (6 out of 8 for galaxies in pairs and 27  of the controls) are LI(N)ERs, which could  in fact be powered by stellar populations instead of a nuclear compact source \citep{Bel16,Hsi17}. 

The aperture corrections are determined for these central SFR measurements   by fitting the photometry of the outer regions of the galaxies \citep{Sal07}.
In the following analysis,  the median SFR and $M_{\ast}$ from the probability distributions for each galaxy are used. Using the average SFR and $M_{\ast}$ does not change the results.

\begin{figure}%
	\centering
	\includegraphics[width=0.47\textwidth]{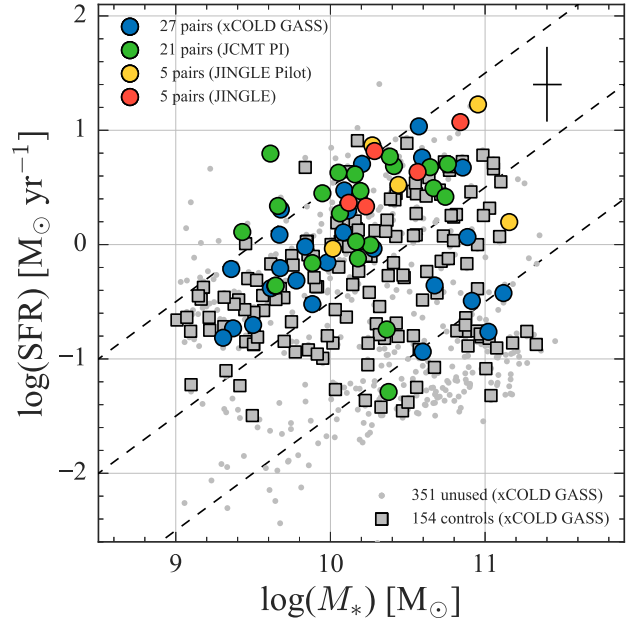}
	\caption{Global star formation rate versus stellar mass relation of galaxies discussed in this work. Galaxies in pairs used in this work are shown as colored symbols. Galaxies from the control sample  are shown as gray squares. Those in the  xCOLD GASS  sample that are not used in the analysis are shown as gray dots (see text for details).  SFR and $M^{\ast}$ values are taken from the MPA-JHU catalog. From top to bottom, the three dashed lines correspond to log(sSFR/yr$^{-1}$) $=$ -9.5, -10.5, and -11.5, respectively.  }%
	\label{fig_SM_SFR_all}%
\end{figure}

The distribution of the pair sample in the SFR versus $M_{\ast}$ plane is shown in Figure \ref{fig_SM_SFR_all} with colored symbols.
Galaxies taken from the control sample drawn from  xCOLD GASS are represented by  gray squares.
Gray dots are the remaining xCOLD GASS targets not selected as control, for reference.

\subsection{Mass Ratio of Pairs}
\label{sec_mr_cal}

Pairs in this work contain both  primary (higher-$M_{\ast}$ galaxy in a pair) and secondary  (lower-$M_{\ast}$ galaxy) galaxies.
For each pair we define the  mass ratio as the stellar mass of the CO observed galaxy divided by that of its companion, and take   the logarithm of the ratio ($\mu$).
A positive value of $\mu$ implies that the observed galaxy is the primary galaxy in the pair, and vice versa. 
A few  companions do not have the MPA-JHU measurement for $M_{\ast}$.
In this case, the mass ratios of these systems are calculated using $M_{\ast}$ from  NASA-Sloan Atlas catalog\footnote{http://nsatlas.org/} for both galaxies.
It is not possible to derive the mass ratio  for the two pairs identified through Galaxy Zoo morphologies as the two galaxies in the pairs are too close to have separated measurements. 
These two galaxies appear to be major mergers, and the CO  observations are made toward the primary galaxies.
They are excluded from the discussion involving mass ratio. 
The distribution of the mass ratio is shown in Figure \ref{fig_merger_props}(c).

\subsection{Gas-phase Metallicity 12$+$log(O/H)}
\label{sec_oh}
In this work, we adopt a metallicity dependent $\alpha_\mathrm{CO}$ (cf. Section \ref{result_absolute_prop}).
 Gas-phase metallicity  is calculated using the O3N2  method empirically calibrated by \cite{Pet04}:
\begin{equation}
12\, +\, \log(\mathrm{O/H})\, =\,8.73 -0.32\mathrm{\left ( \frac{[OIII]}{H\beta }/\frac{[NII]}{H\alpha } \right )}.
\label{eq_OH}
\end{equation}
The emission line fluxes are obtained from the MPA-JHU DR7 release\footnote{https://wwwmpa.mpa-garching.mpg.de/SDSS/DR7/raw\_data.html}.
All the  emission line fluxes of our galaxies in pairs and controls have  S/N higher than 3.5 if the flux uncertainties in the catalog are used, or higher than 2 (mostly $\gg$ 5) if the  scaling factors\footnote{The scaling factors are  2.473, 1.882, 1.566 and 2.039 for H$\alpha$, H$\beta$, [\ion{O}{3}] and [\ion{N}{2}] lines, respectively.} provided in the MPA-JHU DR7 website are applied to the flux uncertainties, which were calculated by the comparisons of the derived line fluxes of galaxies that were observed multiple times. 
Since these emission line  are close in wavelength, dust extinction should have minimal effect on the fluxes.
Galaxies in pairs and controls   have similar ranges of metallicity, from 8.3 to 8.9.
The median values are 8.7 for both populations (Figure \ref{fig_gal_props}(d)).

\section{Results}
\label{sec_results}
\subsection{SFR and Gas Properties in the Full Pair Sample}
\label{result_absolute_prop}
To get an idea of the distribution of galaxy properties within our sample, we first show the normalized distribution of galaxy sSFR, $L_\mathrm{CO}$, and gas properties.
In cases where the CO line is undetected, the 3$\sigma$ upper limits of $L_\mathrm{CO}$ luminosity (\S\ref{sec_control_sample}) are used for the analysis and plots in this paper.

Figure \ref{fig_hists}(a) shows the distribution of  sSFR  in the pair sample as the filled histogram and for the control sample as an open histogram.
While the two distributions peak at the same sSFR, the median sSFR of  galaxies in pairs (log(sSFR/yr$^{-1}$) $=$ -9.9) is higher than that of isolated galaxies (log(sSFR/yr$^{-1}$) $=$ -10.3).
We check whether the distributions in Figure \ref{fig_hists}(a) are sampled from the same parent distribution or not. 
This is tested with the  Kolmogorov-Smirnov (KS) statistic whose results are listed in Table \ref{tab_ks}.
The KS test returns  a $p$-value
$=$ 1.9348 $\times$ 10$^{-5}$, suggesting $>$ 99.9 per cent probability that the sSFR of galaxies in pairs and controls are  two  distinct distributions.
This is in agreement with the many previous studies that have found enhanced star formation in galaxies in pairs \citep[e.g.,][]{Dim07,Eli08,Scu12,Pat13,Kna15}. 
In Sec \ref{sec_rp} and \ref{sec_mr} we return to SFR differences as a function of projected separation and mass ratio, respectively.

Turning now to gas properties, in Figure \ref{fig_hist_Lco}(a), we first show the comparison of integrated $L_\mathrm{CO}$ of galaxies in pairs and controls.
From this figure, it is evident that galaxies in pairs (filled histogram) show higher $L_\mathrm{CO}$ compared to the pool of controls (open histogram).
The difference is significant. The KS test results in a $p$-value of  0.0017.

After having looked  at the observed quantity $L_\mathrm{CO}$, we now derive the physical quantities of gas from the measurements.
The distribution of $M_\mathrm{H_{2}}$ is shown in Figure  \ref{fig_hists}(b).
$L_{\mathrm{CO}}$ is converted to $M_\mathrm{H_{2}}$ by multiplying by the conversion factor, $\alpha _{\mathrm{CO}}$.
The value for  $\alpha _{\mathrm{CO}}$ is calculated following \cite{Acc17}, in which $\alpha _{\mathrm{CO}}$ is  metallicity-dependent, with a second-order dependence on the offset of a galaxy from the star-forming main sequence:
\begin{equation}
\begin{split}
\log \alpha _{\mathrm{CO}}(\pm 0.165)=
& 15.623-1.732[12+\mathrm{(O/H)}] \\
& +0.051\log \Delta (\mathrm{MS}),
\end{split}
\end{equation}
where $\Delta (\mathrm{MS})$ is the distance off the analytical definition of the main sequence by \cite{Whi12}.
No significant difference in the ranges and median values of  $\alpha _{\mathrm{CO}}$  is found  between galaxies in pairs and controls.
The derived $\alpha _{\mathrm{CO}}$ values are in the range of 1.9 -- 10.3 M$_{\odot}$ (K km s$^{-1}$ pc$^{2}$) for the galaxies in pairs (Table \ref{tab_gal}), and  2.1 -- 16.5 M$_{\odot}$ (K km s$^{-1}$ pc$^{2}$) for the controls.
The median $\alpha _{\mathrm{CO}}$ are 3.0 and 3.2 M$_{\odot}$ (K km s$^{-1}$ pc$^{2}$) for galaxies in pairs and controls, respectively.
The   distributions of $M_\mathrm{H_{2}}$ for the galaxies in pairs and controls  largely overlap, but the peak $M_\mathrm{H_{2}}$ of galaxies in pairs is  higher than that of controls.
The median log($M_\mathrm{H_{2}}$/M$_{\odot}$) of galaxies in pairs and controls are 9.3 and 8.9, respectively.
According to the KS test, the differences between galaxies in pairs and controls are real: the probability that their $M_\mathrm{H_{2}}$ are from the same distribution is  less  than 1 per cent ($p$-value $=$ 0.0016).

The  $f_{gas}$ and SFE,  are calculated based on the derived $M_\mathrm{H_{2}}$.
The gas fraction $f_{gas}$ is defined as
\begin{equation}
f_{gas}= \frac{M_\mathrm{H_{2}}}{M_\mathrm{H_{2}}+M_{\ast }}.
\end{equation}
Note that some studies adopt the definition of $f_{gas}= M_\mathrm{H_{2}}/M_{\ast }$.
The two definitions are approximately the same when $M_\mathrm{H_{2}}$ $\ll$ $M_{\ast }$ (i.e., low $f_{gas}$).
Our results would be qualitatively the same if we adopted this definition.
Figure \ref{fig_hists}(c) shows the normalized distribution of molecular gas fraction.
The log($f_{gas}$) of controls (open histogram) spans the range -2.37 -- -0.64 (corresponding to 0.4 -- 22.9\%)  with a median value of -1.26 (5.5\%). 
The values are  consistent with other studies of  galaxies  in the local Universe \citep[e.g.,][]{Ler08,Bot09,Sai17}.
On the other hand, the range of log($f_{gas}$) of galaxies in pairs (filled histogram)  shifts to larger values from -1.81 to -0.20 (1.5 -- 63.1\%),  with a median value of -1.03 (9.3\%).
A KS test of $f_{gas}$  yields a $p$-value $=$ 3.9535 $\times$ 10$^{-7}$.
There is less than a 0.1 per cent chance that the two samples come from the same distribution.

The SFE is defined as:
\begin{equation}
\frac{\mathrm{SFE}}{[\mathrm{yr^{-1}}]}  = \frac{\mathrm{SFR}}{M_{\mathrm{H_{2}}}}
\end{equation}
and is shown in Figure \ref{fig_hists}(d). 
The ranges and peaks of SFE appear closely matched for galaxies in pairs and controls.
Median log(SFE/yr$^{-1}$) are -9.02 and -9.13 for galaxies in pairs and controls, respectively.
The values of SFE are comparable with other studies of nearby isolated galaxies \citep[e.g.,][]{Ler08,Koy17}.
A KS test suggests that  the two samples are drawn from the same distribution ($p$-value $=$ 0.2128).

\begin{table}[]
	\centering
\caption{Summary of the Kolmogorov-Smirnov (KS) test results for galaxies in pairs and controls. }
\label{tab_ks}
	\begin{tabular}{p{1.7cm}p{2.6cm}p{2.6cm}}
		\hline
		\hline
                   & raw value            & offset value ($\Delta$)   \\
& (\S\ref{result_absolute_prop})       & (\S\ref{sec_offset})       \\
\hline
(s)SFR             & 1.9348 $\times$ 10$^{-5}$ & 8.2130 $\times$ 10$^{-6}$ \\
$L_\mathrm{CO}$    & 0.0017                    & 4.3003 $\times$ 10$^{-5}$ \\
$M_\mathrm{H_{2}}$ & 0.0016                    & 1.1890 $\times$ 10$^{-6}$ \\
$f_{gas}$          & 3.9535 $\times$ 10$^{-7}$ & 8.7772 $\times$ 10$^{-7}$ \\
SFE                & 0.2128                    & 0.4806   \\
		\hline      
	\end{tabular}
\end{table}

We have shown in this section that there are statistically significant differences in SFR, $M_\mathrm{H_{2}}$ and $f_{gas}$ between the galaxies in pairs and the full control sample, and that there is no significant difference in SFE between these two populations.  
However, we have not yet considered possible differences in the fundamental properties of these two samples, such as stellar mass and redshift.  Since SFR and gas properties themselves depend on these underlying characteristics, a rigorous comparison requires a careful matching between the galaxies in pairs and control sample.  We investigate this in the following section.

\subsection{Offset of SFR and Gas Properties}
\label{sec_offset}
\begin{figure*}%
	\centering
		\includegraphics[width=0.9\textwidth]{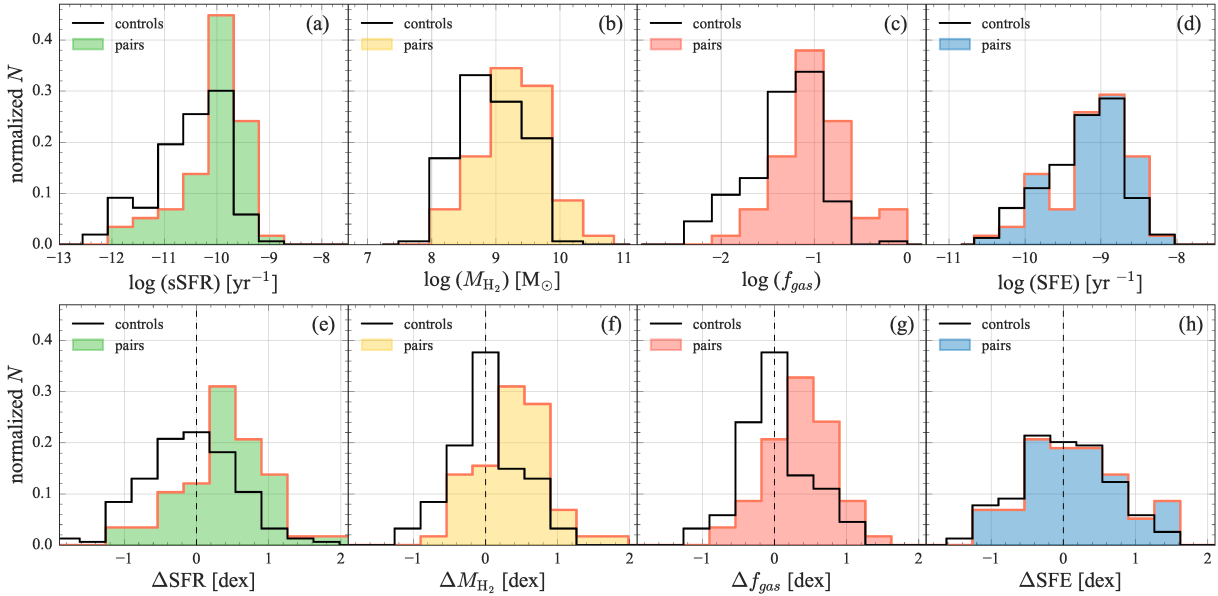}
	\caption{Histograms showing the distribution of physical quantities sSFR, $M_\mathrm{H_{2}}$, $f_{gas}$, and SFE in upper row, and the offset of these properties with respect to the control sample in the lower row.
	 The galaxies in pairs and controls  are plotted as filled and open histograms, respectively. The vertical dashed lines  in the lower panels indicate zero enhancement. The enhancements of SFR,  $M_\mathrm{H_{2}}$ and $f_{gas}$ are observed statistically significant  for both raw and offset quantities (Table \ref{tab_ks}). The strength of SFE offset is not as large as that of other properties, and  a  Kolmogorov-Smirnov test suggests that the difference is not significant.
		}%
	\label{fig_hists}%
\end{figure*}

\begin{figure}%
	\centering
	\includegraphics[width=0.45\textwidth]{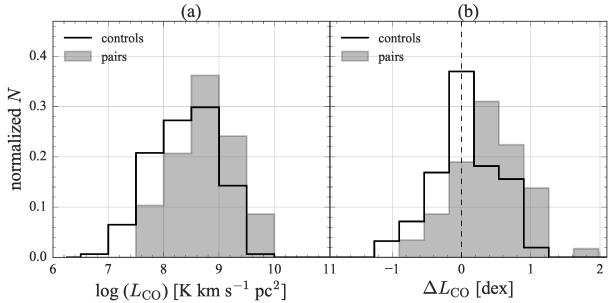}
	\caption{Distributions of $L_\mathrm{CO}$ (a) and $\Delta$$L_\mathrm{CO}$ (b). The galaxies in pairs and controls  are plotted as filled and open histograms, respectively. The vertical dashed line  in panel (b) indicates zero enhancement. The figures confirm that the enhancement of $M_\mathrm{H_{2}}$ and $f_{gas}$ of gas in galaxies in pairs are already indicated from the comparison from the observed    quantity $L_\mathrm{CO}$, i.e., before the conversion to $M_\mathrm{H_{2}}$. 
	}%
	\label{fig_hist_Lco}%
\end{figure}

In order to fairly compare the properties of the galaxies in pairs and controls on a galaxy-by-galaxy basis, and accounting for dependences on properties such as redshift and $M_{\ast}$, we compute in this section ``offset''  quantities.
Our approach follows closely that of \cite{Vio18} for H$_{2}$ fractions in a smaller sample of galaxies in pairs, and also that of \cite{Ell18} for \ion{H}{1} fractions in post-mergers.
Each  galaxy, including  pairs and controls,  is matched in redshift, stellar mass, and effective radius with a minimum of five control galaxies  from the pool of controls.
The  tolerance of stellar mass, redshift, and effective radius are 0.1 dex, 0.005, and 25\%, respectively.
The criteria are allowed to grow by 0.1 dex, 0.005, and 5\% respectively, until the minimum required number of control galaxies  is reached. Most of the galaxies could find sufficient control galaxies in the first round.
The ``offset'' of a galaxy property (P) is defined as,
\begin{equation}
\Delta (\mathrm{P})=\log(\mathrm{P_{gal}})-\log(\mathrm{median(P_{control})}),
\end{equation}
where $\mathrm{P_{gal}}$ is the property of the galaxy in question and median(P$_\mathrm{control}$) is the median property of its  control galaxies.
We should emphasize that as this is taken in the logarithm form, it really is a ratio of a value of galaxy in question against the median value of its controls.
A positive offset  represents an \emph{enhancement}, while  negative value implies a \emph{suppression} of the property.
Distributions of  offset properties  are presented in Figure \ref{fig_hist_Lco}(b) for $L_\mathrm{CO}$ and Figure \ref{fig_hists}(e)--(h) for other physical quantities related to SFR\footnote{In Section \ref{result_absolute_prop}, we use ``sSFR'' to express the  absolute value of star formation rate so that the dependence of SFR on $M_{\ast}$ is considered. The offset property of star formation is defined as the difference of  ``SFR'' between galaxies in pairs and controls. The two star formation rates are not different, as the $M_{\ast}$  is matched when selecting control sample to calculate the star formation rate offset.} and gas properties. 
Open and filled histograms denote the offset properties of controls and galaxies in pairs, respectively.
The median values for the controls are around zero, as expected, the width showing the intrinsic spread of the quantities plotted. 

The enhancement of  SFR is still present and significant ($p$-value $=$ 8.2130 $\times$ 10$^{-6}$, see Table \ref{tab_ks}), as indicated by the peaks of the distributions (Figure \ref{fig_hists}(e)).
The median $\Delta$SFR is 0.40 dex (a factor of $\sim2.5$ enhancement),  confirming the well-known fact that, statistically,    galaxy-galaxy interaction enhances the SFR, but not dramatically so; moreover, there is a large spread in the enhancement of SFR \citep[e.g.,][]{Scu12,Kna15}.

Figure \ref{fig_hist_Lco}(b) confirms that the enhancement of the amount of gas in galaxies in pairs is already indicated from the comparison from the observed $L_\mathrm{CO}$, i.e., before the conversion to $M_\mathrm{H_{2}}$. 
The KS test yields a $p$-value of 4.3003 $\times$ 10$^{-5}$.
The median $\Delta$$L_\mathrm{CO}$ of galaxies in pairs is 0.40 dex.

The median $\Delta M_\mathrm{H_{2}}$ and $\Delta f_{gas}$ are enhanced by  similar factors, about 0.37 and 0.44 dex, respectively (Figure \ref{fig_hists}(f) and (g)).
We also apply the KS test to the  offset distributions of galaxies in pairs and controls.
For  $M_\mathrm{H_{2}}$ and $f_{gas}$, the KS test gives a $<$ 1\% probability of the two distributions being drawn from the same parent distribution (Table \ref{tab_ks}).

The strength of SFE offset is not as large as that of other properties.
The median $\Delta$SFE implies an offset by 0.14 dex with respect to the controls. 
However, a KS test suggests that the difference is not significant;  there is a high probability that the two distributions (control and pairs) are drawn from the same population.  

\begin{figure*}%
	\centering
	\includegraphics[width=0.8\textwidth]{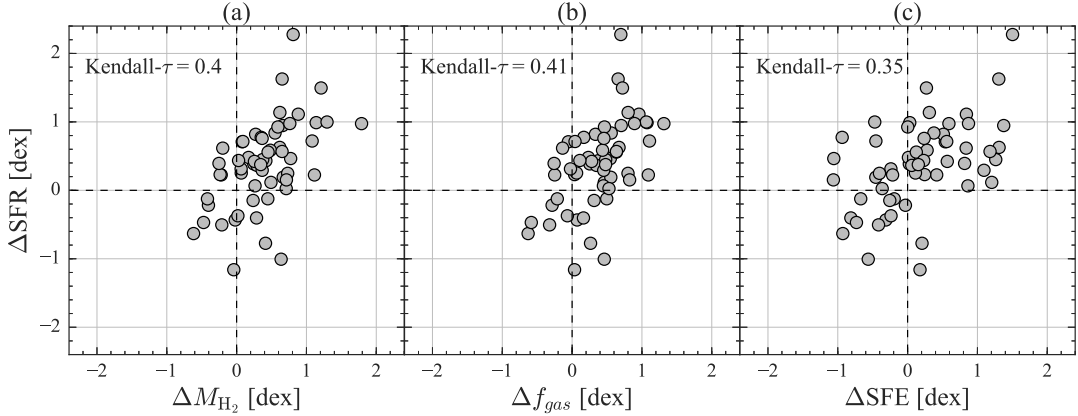}
	\caption{$\Delta$SFR versus $\Delta M_\mathrm{H_{2}}$(a),  $\Delta f_{gas}$ (b), and $\Delta$SFE (c) of the galaxies in pairs.  The dashed lines   indicate zero enhancement. The Kendall's tau correlation coefficient  between two variables are given in the upper-left corner of each panel. The figure indicates that all molecular gas properties ($M_\mathrm{H_{2}}$, $f_{gas}$, and SFE) are expected to influence  SFR.
	}%
	\label{fig_scatter2_all_xcoo}%
\end{figure*}

In order to investigate which gas property is most strongly correlated with SFR enhancement, in  Figure \ref{fig_scatter2_all_xcoo} we plot  $\Delta f_{gas}$, $\Delta M_\mathrm{H_{2}}$ and $\Delta$SFE versus $\Delta$SFR.
All gas property offsets increase with $\Delta$SFR.
We use Kendall's $\tau$ correlation coefficient to quantify the strength of the dependence.
The computation yields positive correlations of  0.40, 0.39 and 0.35 for $\Delta M_\mathrm{H_{2}}$, $\Delta f_{gas}$ and  $\Delta$SFE, respectively.
Based on this  figure, we can speculate that all molecular gas properties ($M_\mathrm{H_{2}}$, $f_{gas}$, and SFE) are expected to influence  SFR.

\begin{figure}%
	\centering
	\includegraphics[width=0.47\textwidth]{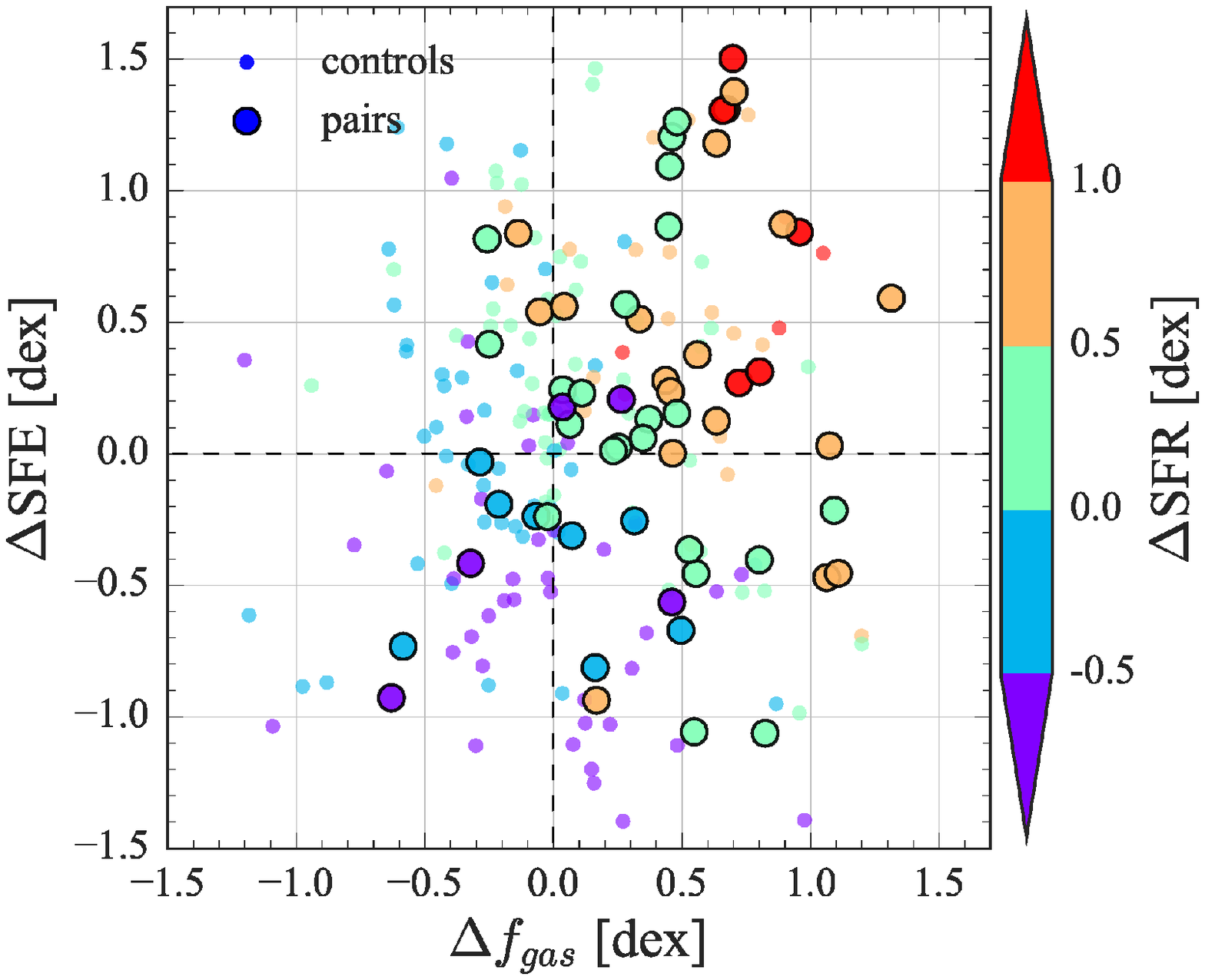}
	\caption{$\Delta$SFE versus $\Delta f_{gas}$, color coded by $\Delta$SFR. The large and small circles represent galaxies in pairs and controls respectively. The highest $\Delta$SFR ($>$ 0.5 dex) predominantly occur in galaxies with both  enhanced $\Delta f_{gas}$ \emph{and} enhanced $\Delta$SFE, however,  enhanced $\Delta f_{gas}$ and $\Delta$SFE together do not always result in  high magnitude of  $\Delta$SFR (but it is true that almost all galaxies with positive values of $\Delta f_{gas}$ and $\Delta$SFE show enhanced SFR). Galaxies may not have enhanced SFR if only $f_{gas}$ or SFE is enhanced (upper left and lower right corners of the figure). Finally, galaxies associated with both suppressed SFE and $f_{gas}$ are likely to have suppressed SFR as well (lower left corner of the figure).}%
	\label{fig_scatter4_all_xcoo}%
\end{figure}

Figure \ref{fig_scatter4_all_xcoo} presents the $\Delta$SFE versus $\Delta f_{gas}$, color coded by $\Delta$SFR.
The large and small circles represent galaxies in pairs and controls respectively.
Various inferences can be drawn from this figure.
The highest $\Delta$SFR predominantly occur in galaxies with both  enhanced $\Delta f_{gas}$ and enhanced $\Delta$SFE, however,  enhanced $\Delta f_{gas}$ and $\Delta$SFE together do not always result in  high magnitude of  $\Delta$SFR (but almost all galaxies with positive values of $\Delta f_{gas}$ and $\Delta$SFE show enhanced SFR).
Galaxies may not have enhanced SFR if only $f_{gas}$ or SFE is enhanced.
Finally, galaxies associated with both suppressed SFE and $f_{gas}$ are likely to have suppressed SFR as well.

\subsection{$\Delta$SFR, $\Delta$$f_{gas}$, and $\Delta$SFE as a Function of Projected Separation }
\label{sec_rp}

\begin{figure}%
	\includegraphics[width=0.45\textwidth]{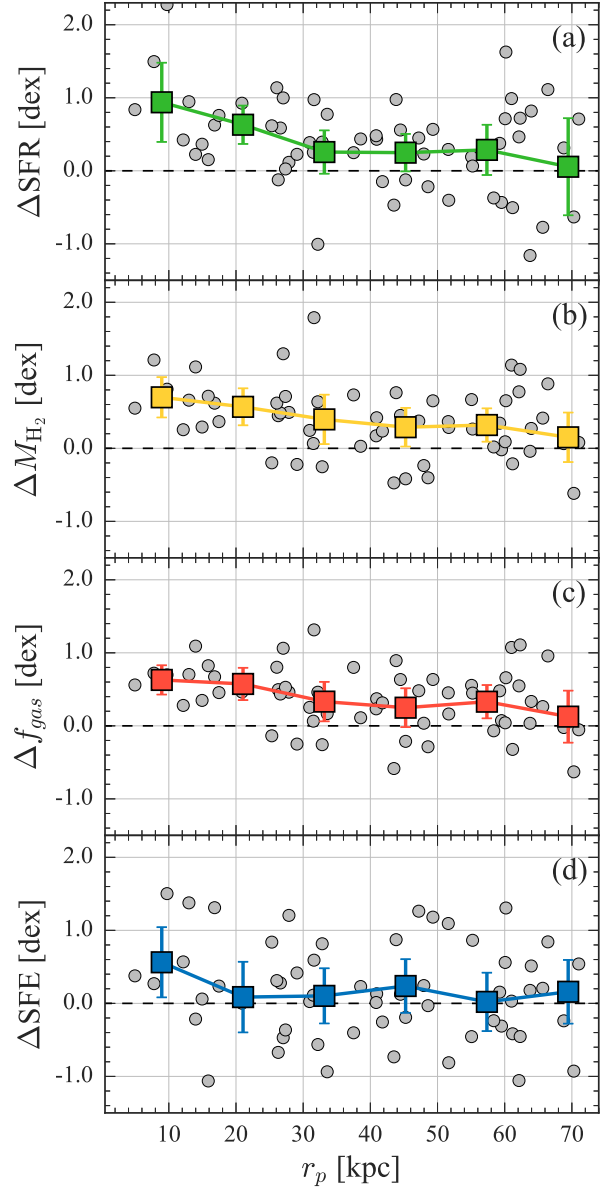}
	\caption{Offset properties as a function of projected galaxy separation  for our sample. Gray circles denote individual galaxies. Mean values per $r_{p}$ are indicated with colored squares. Error bars are obtained by calculating the sample standard deviation and dividing by $\sqrt{N}$, where $N$ is the number of galaxies at each $r_{p}$ bin. The horizontal lines indicate no enhancement. $\Delta$SFR, $\Delta$$M_\mathrm{H_{2}}$ , and $\Delta$$f_\mathrm{gas}$ all increase with decreasing pair separation  over the range from $\sim$ 70 to 10 kpc. However, any SFE enhancement is only significant at the smallest pair separations.}%
	\label{fig_grad_all_xco}%
\end{figure}

Figure \ref{fig_grad_all_xco}(a) presents  $\Delta$SFR of our galaxies in pairs as a function of $r_{p}$.
Gray circles denote individual galaxies in pairs, colored squares show the mean $\Delta$SFR at different $r_{p}$ bins.
Error bars are obtained by calculating the sample standard
deviation and dividing by $\sqrt{N}$, where $N$ is the number of galaxies at each $r_{p}$ bin.
The dashed horizontal line denotes zero enhancement.
At small separations, galaxies in pairs are found to have substantially higher SFR than their controls, with  median $\Delta$SFR increasing from $\approx$ 0.2 dex at $\sim$ 70 kpc to $\sim$ 0.9 dex at $<$ 10 kpc.
The fastest rise of $\Delta$SFR occurs at $r_{p}$ $\approx$ 20 -- 30 kpc, as also shown in \cite{Scu12} and \cite{Pat13} with much larger sample sizes than this work, and in  \cite{Bus17} with cosmological simulations.
The $\Delta$SFR (0.9 $\pm$ 0.5 dex) in the lowest $r_{p}$  bin is higher than that of post-mergers (0.5 $\pm$ 0.1 dex) in \cite{Ell13}.
This may be due to a mix of  post-mergers that are already quenched and  that are forming stars actively, as the SFR and duration of enhanced SFR at the coalescence phase depends on various merger configurations  \cite[e.g.,][]{Dim07,Dim08,Bus17,Tho18}.

The enhanced SFR as two galaxies approach each other could be interpreted as the direct evidence of tidally triggered star formation.
\cite{Par09} investigate the dependence of galaxy properties on both the small- and large-scale environments.
They find that galaxy properties, such as H$\alpha$ equivalent width, surface brightness profile, and colors,  abruptly change when $r_{p}$ corresponds to the 0.05 $\times$ virial  radius  of  the nearest  neighbor  galaxy (see their Figure 6 and 7).
This corresponds to $\sim$ 20 kpc for the galaxies in their sample.
This interpretation can be applied to our result of the boost of $\Delta$SFR at $r_{p}$ $\approx$ 20 -- 30 kpc.
However we should note that  this characteristic radius depends on the stellar mass of the sample galaxies because the hydrodynamic interactions between galaxies depend on the stellar mass \citep{Par09}.

It should be noted that there is significant scatter within each $r_{p}$ bin \citep[see also][]{Scu12}.
There are several reasons for the scatter.
Firstly, the peak in SFR enhancement does not always occur near coalescence.
SFR could  reach the peak when two galaxies are still several tens of kpc apart  \citep{Dim08,Spa16}. 
Moreover, $\Delta$SFR depends on the mass ratio.
This will be discussed in Section \ref{sec_mr}.
Many studies have stressed the importance of the properties of the companion in determining the SFR enhancement \citep{Par09,Hwa10,Xu12,Cao16}.
Specifically, the SFR of spirals in spiral-spiral pairs are more likely to be  enhanced compared to the spirals in mixed spiral-elliptical pairs.
The suppression (or  zero enhancement) of star formation in the disks in the mixed  pairs  may be caused by the extended X-ray halos (i.e., hot gas) of an early type companion of a  spiral galaxy and  prevent the spiral from forming stars, or there is no inflow of cold gas from the early-type companion \citep{Par09,Hwa10}.
Furthermore, SFR enhancement seems to be correlated with the properties of the orbit of the two interacting galaxies as shown by simulations of \cite{Spa16}, in which  high-density  gas  preferentially appears in head-on mergers with very high collision velocities. 
This scenario is difficult to test directly by observation due to the ambiguity, even when detailed models can be constructed, in   a system's geometry and orbital parameters.
Finally, SFR enhancement is also found to correlate with HI fraction  \cite[e.g.,][]{Scu15}.
Since galaxies in pairs used in this work  are not restricted to any specific merger property, the scatter in each $r_{p}$ bin is somewhat expected.
We should note that part of this scatter is also due to the fact that the $r_{p}$ is not a direct measure of the merging sequence because galaxies in pairs would merge after several encounters and their orbital geometry is complicated. 
This presumably also introduces some extra scatter in the relation.

Turning to the gas properties, panels (b) and (c) present the change of $\Delta M_\mathrm{H_{2}}$ and $\Delta f_{gas}$ with $r_{p}$, respectively. 
All symbols are as defined for panel (a). 
$\Delta M_\mathrm{H_{2}}$ and $\Delta f_{gas}$ versus $r_{p}$ show very similar behavior to $\Delta$SFR. 
$\Delta M_\mathrm{H_{2}}$ and $\Delta f_{gas}$ gradually increase from $\sim$ 0 dex at 70 kpc to $\sim$ 0.7 and 0.6 dex respectively at $<$ 10 kpc. 
In particular, the figures show that  almost all close pairs ($r_{p}$ $\leq$ 25 kpc)  appear to have $M_\mathrm{H_{2}}$ and $f_{gas}$ enhancements.

Here we compare our results of $\Delta f_{gas}$ with other studies  in which the offset of gas properties are also calculated.
The 11 galaxies in pairs in \cite{Vio18} have a $r_{p}$ range of 16 -- 30 kpc. 
The median offset $M_\mathrm{H_{2}}$ and $f_{gas}$ of their  galaxies in pairs are 0.34 and 0.40 dex, respectively.
For our galaxies in pairs in the same $r_{p}$ range, the median $\Delta M_\mathrm{H_{2}}$ and $\Delta f_{gas}$ are 0.49 and 0.46, respectively, slightly higher than that of \cite{Vio18}.
It may  simply be due to low number statistics.
The degree of the  $f_{gas}$ enhancement of our galaxies in pairs at short  $r_{p}$ ($\sim$ 0.6 dex)   is consistent to that of the sample of post-mergers  (Sargent et al. in prep.).

The dependence of $\Delta$SFE on $r_{p}$ is  different from that of $\Delta$SFR, $\Delta$$M_\mathrm{H_{2}}$ and $\Delta$$f_{gas}$. 
Statistically, SFE enhancements only occur  at the smallest pair separations ($r_{p}$ $<$ 20 kpc) by $\sim$0.5 dex.
In other words, although overall the $\Delta$SFE shows zero enhancement (Figure \ref{fig_hists}(h)), but there is a systematic offset for the smallest $r_{p}$ galaxies.
The scatter of $\Delta$SFE is large at large $r_{p}$.
We will discuss the possible contribution of the large scatter later in the discussion section (Section \ref{sec_what_drives}).

The statistical significances of the correlations  are assessed by calculating the Kendall's $\tau$ correlation coefficients.
The  correlation coefficients are -0.33, -0.29, and -0.29 for $r_{p}$ versus $\Delta$SFR, $\Delta$$M_\mathrm{H_{2}}$, and $\Delta$$f_{gas}$, respectively.
If we restrict the analysis to  galaxies with $r_{p}$ $<$ 30 kpc where the offset values appear to rise more rapidly with decreasing  $r_{p}$, the  correlation coefficients become -0.40, -0.30, and -0.31 for $\Delta$SFR, $\Delta$$M_\mathrm{H_{2}}$, and $\Delta$$f_{gas}$, respectively.
The correlation coefficients suggest that there are only  marginal anti-correlations between $\Delta$SFR, $\Delta$$M_\mathrm{H_{2}}$, and $\Delta$$f_{gas}$ and the pair separation, possibly due to the large scatter at a fixed $r_{p}$.
The absence of a correlation between  $r_{p}$ and $\Delta$SFE is also suggested by the correlation coefficient of -0.10 across all galaxies in pairs. 
The correlation coefficient of $r_{p}$ and $\Delta$SFE becomes -0.22 for galaxies with $r_{p}$ $<$ 30 kpc.

\subsection{$\Delta$SFR, $\Delta$$f_{gas}$, and $\Delta$SFE as a Function of Mass Ratio }
\label{sec_mr}

\begin{figure}%
	\includegraphics[width=0.45\textwidth]{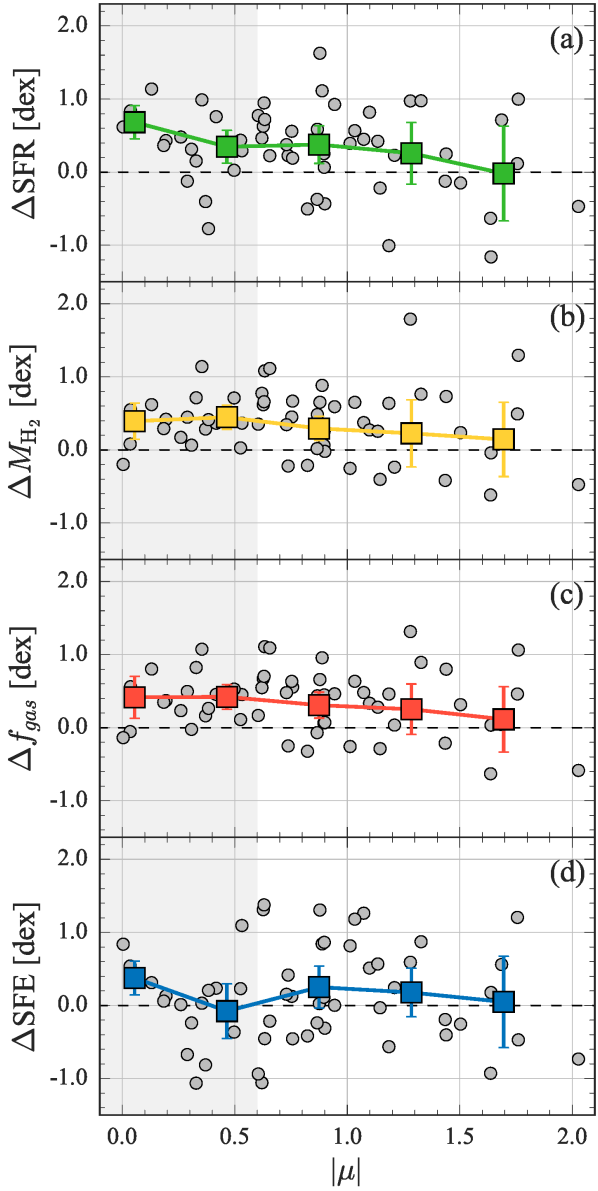}
	\caption{Offset properties as a function of the absolute value of stellar mass ratio $\left | \mu  \right |$ of the galaxies in pairs. The  major merger regime ($\left | \mu  \right |$ $<$ 0.6) is highlighted in  gray. The individual galaxies in pairs are shown with gray circles and the means are in colored symbols. $\Delta$SFR, $\Delta$$M_\mathrm{H_{2}}$ , and $\Delta$$f_\mathrm{gas}$  exhibit a  trend with  mass ratio of the two galaxies in a pair. We find no apparent dependence between the mass ratio and $\Delta$SFE. Any SFE enhancement is only significant in the equal-mass pairs ($\left | \mu  \right |$ $\approx$ 0).}%
	\label{fig_massratio_all_xco}%
\end{figure}

Our sample covers about two orders of magnitude  mass ratio, and includes both primary and secondary galaxies.
Since the number of secondary galaxies is considerably smaller than the primary galaxies (11 versus 45), not allowing us to compare between  these two populations, in this section we consider the absolute value of mass ratio $\left | \mu  \right |$.
Figure \ref{fig_massratio_all_xco}(a) presents $\Delta$SFR as a function of $\left | \mu  \right |$. 
The  major merger regime ($\left | \mu  \right |$ $<$ 0.6) is colored in  gray.
The individual galaxies in pairs are shown with gray circles and the means are in colored symbols.

The mean $\Delta$SFRs are progressively higher for smaller $\left | \mu  \right |$ values.
Most, but not all,    major mergers in our sample show SFR enhancement. 
As suggested by simulations, a major merger is not   inevitably accompanied by significant SFR enhancement, depending on the geometry of the collisions \citep{Cox08,Mor15,Spa16}.
This may explain why some of the   major mergers show low $\Delta$SFR.

$\Delta M_\mathrm{H_{2}}$ and $\Delta f_{gas}$ exhibit a similar trend as $\Delta$SFR in Figure  \ref{fig_massratio_all_xco}(b) and (c), increasing   from large to low mass ratio.
On the other hand, the $\Delta$SFE  trend with mass ratio  is not as strong as for other properties. 
There is no significant  difference in $\Delta$SFE  across $\left | \mu  \right |$. 
Statistically, SFE enhancements only occur  in the equal-mass pairs ($\left | \mu  \right |$ $\approx$ 0) by $\sim$0.4 dex.

We also quantify  the degree of correlations between the offset properties and $\left | \mu  \right |$ using the Kendall's $\tau$ correlation coefficient. 
The correlation coefficients are -0.25, -0.18, and -0.20, and -0.001 for $\Delta$SFR, $\Delta M_\mathrm{H_{2}}$, $\Delta f_{gas}$, and $\Delta$SFE, respectively, indicating marginal trends for equal-mass pairs to have  higher $\Delta$SFR, $\Delta M_\mathrm{H_{2}}$, $\Delta f_{gas}$, but not for $\Delta$SFE.

\section{Discussion}
\label{sec_discuss}
\subsection{Are the Enhanced  $\Delta M_\mathrm{H_{2}}$ and $\Delta f_{gas}$ Real?}
\label{sec_real}
While several studies have suggested the enhancement of $\Delta M_\mathrm{H_{2}}$ and $\Delta f_{gas}$ in galaxies in pairs \citep[e.g.,][]{Com94,Cas04,Vio18}, it  remains unclear from where this mass excess originates.
We first check whether the enhanced  $\Delta M_\mathrm{H_{2}}$ and $\Delta f_{gas}$ are real. 

The enhanced  $\Delta M_\mathrm{H_{2}}$ and $\Delta f_{gas}$ could be the contamination from the CO emission of the companions.
We check  the possible contamination by comparing the projected separation between two galaxies in a pair and the beam size (radius).
For   galaxies  with $r_{p}$ $>$ 30 kpc, the distance to the companion is well beyond  the beam area, with a median distance of 8 $\times$  beam radius.
For   galaxies with 11 $<$ $r_{p}$ $<$ 30 kpc, companions are located at  1.5 to 10 $\times$  beam radius away from the pointing of the CO observations; the median distance is 3 $\times$  beam radius. 
For  the  close pairs  with  $<$ 11 kpc,  the beam areas cover a part  of the disk of  their companions.
In one of the cases, the nucleus of the companion  falls within the beam area. 
Consequently, there must be a non-negligible contribution of CO emission from the companions in these measurements.
However, it is unlikely the sole cause of the enhanced $\Delta M_\mathrm{H_{2}}$ and $\Delta f_{gas}$ as the offsets are considerably larger than a factor of 2 (or 0.3 dex) and the  positive offsets are seen in galaxies beyond the the separation for contamination.
Moreover,  $\Delta f_{gas}$ enhancement is seen in post-mergers by $\sim$ 0.6 dex (Sargent et al. in prep.), where there is  only one galaxy in the beam.
Therefore, an additional mechanism
must be involved in boosting the amount of molecular gas.
Nonetheless, we conservatively quote  an overestimation (underestimation) of 0.3 dex, corresponding to $\sim$ 2 times, for the derived $\Delta M_\mathrm{H_{2}}$ and $\Delta f_{gas}$ ($\Delta$SFE), assuming a situation consisting of a major merger where both galaxies fall within the beam area.

The observed enhanced $\Delta M_\mathrm{H_{2}}$ and $\Delta f_{gas}$  could partially be a selection effect towards high SFR objects (and thus  likely molecular gas rich) since  some of the pair sample (PI program, JINGLE and JINGLE Pilot) are selected based  reaching a minimum signal to noise level of 4 in an on-source time shorter than 250 min. which is estimated on the basis of observed SFR and the empirical SFR -- $M_\mathrm{H_{2}}$ relation (\S\ref{sec_data}).
We cannot fully rule out the possibility that with deeper data we would detect more galaxies in pairs with $\Delta M_\mathrm{H_{2}}$ and $\Delta f_{gas}$ comparable to that of the control sample.
 For example,  a far-infrared ($\propto$ SFR) selection of galaxies in the Coma cluster prevented \cite{Cas91} to  find molecular gas deficient galaxies in clusters, while H$_{2}$-deficient galaxies are now widely found when other selection criteria are used such as stellar mass \citep[e.g.,][]{Bos14a}.

\begin{figure*}
	\centering
	\begin{subfigure}
		\centering
	\includegraphics[width=0.95\textwidth]{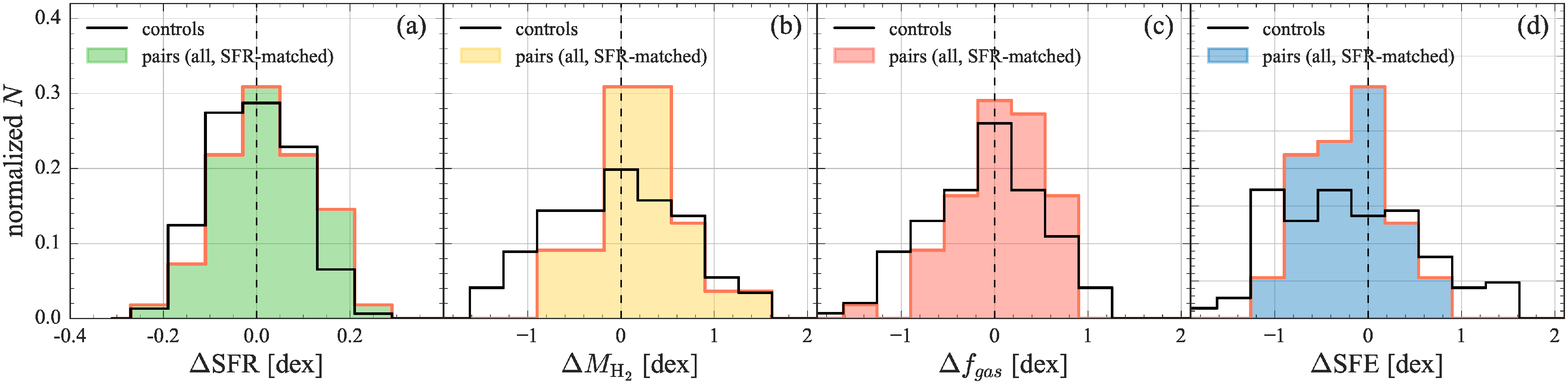}
		\label{fig_hist_delta_gas_prop_strmatch_xco}
	\end{subfigure}
	\begin{subfigure}
	\centering
	\includegraphics[width=0.95\textwidth]{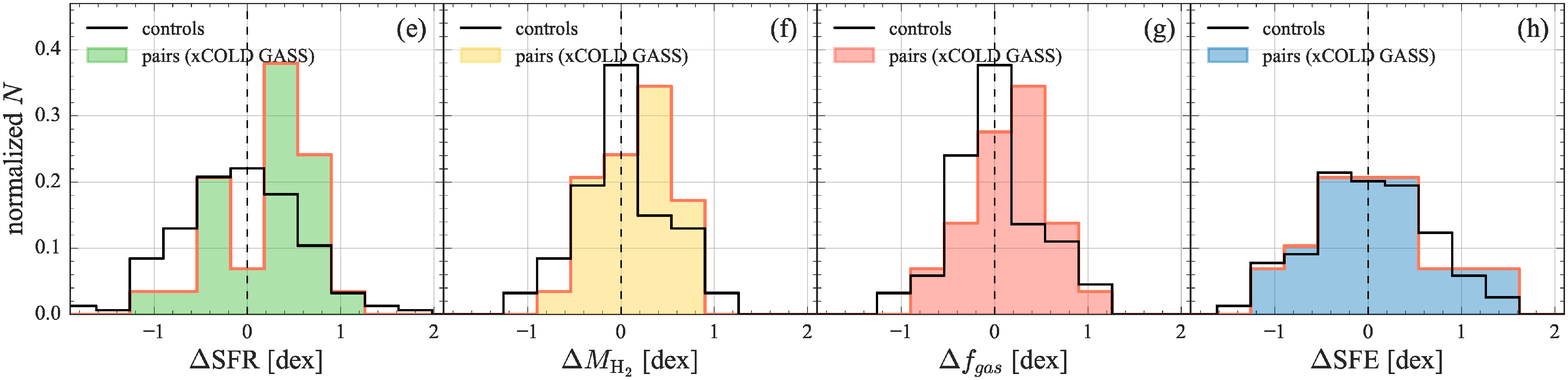}
	\label{fig_hist_delta_gas_prop_xcg_xco}
\end{subfigure}%
	\caption{(a) -- (d): Normalized distribution of $\Delta$SFR, $\Delta M_\mathrm{H_{2}}$, $\Delta f_{gas}$, and  $\Delta$SFE. The plot is similar to Figure \ref{fig_hists}(e)--(h), but we additionally match the galaxies and control sample in SFR. Therefore the values along the $x$-axis in panel (a) are considerably smaller than in the other panels. Galaxies in pairs with high sSFR  are not considered here (see text for details) due to the lack of suitable controls.  With SFR as one of matching parameters, the enhancements of $\Delta M_\mathrm{H_{2}}$ and $\Delta f_{gas}$ are still present in galaxies in pairs.    SFE seems to be suppressed in galaxies in pairs after matching in SFR. This is because that SFE is SFR  divided by $M_\mathrm{H_{2}}$  by definition. (e) -- (h): The plots are made using the data values that generated Figure \ref{fig_hists}(e)--(h) (without matching in SFR), but only galaxies in pairs from  xCOLD GASS are shown. Since the control galaxies are drawn from  xCOLD GASS as well,  any potential bias induced by  different selection criteria for galaxies in pairs (Section \ref{sec_data})  and any uncertainty in the conversion between CO(2-1) to CO(1-0) intensities are removed. It is evident that the distributions of $\Delta$SFR, $\Delta M_\mathrm{H_{2}}$, $\Delta f_{gas}$ still peak at higher values with respect to the controls when only the xCOLD GASS galaxies in pairs are considered.  }
	\label{fig_hist_tests}
\end{figure*}

In order to consider such effect,  we additionally match the galaxies in pairs and controls in SFR with an initial tolerance of 0.1 dex and a step of  0.05 dex.
In carrying out this test with extra matching parameter, we find that it is difficult  to have  at least five control galaxies for each pair, due to the limited number of  pool galaxies, thus the minimal number of control galaxies is  instead set to three.
Moreover,  we were unable to identify control galaxies for some galaxies in pairs with high  sSFR; and these objects are excluded from the analysis here.  The results are presented in Figure \ref{fig_hist_tests}(a)--(d).

With SFR as one of matching parameters (Figure \ref{fig_hist_tests}(a)), the enhancements of $\Delta M_\mathrm{H_{2}}$ and $\Delta f_{gas}$ are still present in galaxies in pairs, although with smaller values.
The distribution of $\Delta M_\mathrm{H_{2}}$ and $\Delta f_{gas}$ when additionally matched in SFR are presented in Figure \ref{fig_hist_tests}(b) and (c). 
The median $\Delta M_\mathrm{H_{2}}$ and $\Delta f_{gas}$ become  0.19  and 0.11  dex ($\sim$ 55\% and 30\%), respectively.
This reduction of the difference in molecular gas content between galaxies in pairs and controls when they additionally matched in SFR is  also found by \cite{Vio18}.
However, we note that the  median values here should be treated as lower limits since the galaxies in pairs with high sSFR are not considered,  and these galaxies potentially have large $\Delta M_\mathrm{H_{2}}$ and $\Delta f_{gas}$.

After matching in SFR, SFE seems to be suppressed in galaxies in pairs. 
The median $\Delta$SFE is -0.20 dex (Figure \ref{fig_hist_tests}(d)). 
This is not unexpected, since SFE is SFR (matched with controls) divided by $M_\mathrm{H_{2}}$ (enhanced) by definition.

In addition, we also  perform another test to examine the potential bias induced by  different selection criteria for galaxies in pairs (cf. Section \ref{sec_data}), that is, only including galaxies from  xCOLD GASS (without matching in SFR).
In this case, the selection of a control sample uses exactly the same criteria as for  galaxies in pairs.
This would also remove any uncertainty in the conversion between CO(2-1) to CO(1-0) intensities.
The results are shown in Figure \ref{fig_hist_tests}(e)--(h) (specifically, these plots are made with  the same data values that generated  Figure \ref{fig_hists}(e)--(h), but only galaxies in pairs from  xCOLD GASS are shown).
It is evident that  the distributions of $\Delta$SFR, $\Delta M_\mathrm{H_{2}}$ and $\Delta f_{gas}$ still peak at higher values with respect to the controls when only the xCOLD GASS galaxies in pairs are considered. 
The median offset values are $\sim$ 0.3 dex for $\Delta$SFR, $\Delta M_\mathrm{H_{2}}$ and $\Delta f_{gas}$, and $\sim$ 0.1 dex for $\Delta$SFE.

Overall,   we  argue that the contamination from companions and selection effect are not the main cause of the enhanced $\Delta M_\mathrm{H_{2}}$ and $\Delta f_{gas}$.

\subsection{What drives the gas reservoir enhancement?}
\label{sec_what_drives}

The  main results of this work are: (1) there is a clear enhancement in the gas reservoir ($M_\mathrm{H_{2}}$ and $f_{gas}$) in galaxies in pairs  (\S\ref{sec_offset}), (2)  $M_\mathrm{H_{2}}$ and $f_{gas}$ increase by a similar degree to SFR, while the SFE is compatible with not being enhanced  (\S\ref{sec_offset}), (3) the correlation with $\Delta$SFR is stronger for $\Delta M_\mathrm{H_{2}}$ and $\Delta f_{gas}$ than $\Delta$SFE (\S\ref{sec_offset}), and (4) the dependences of $\Delta M_\mathrm{H_{2}}$ and $\Delta f_{gas}$ on merger configurations are similar to that of $\Delta$SFR, with  $\Delta$SFR, $\Delta M_\mathrm{H_{2}}$ and $\Delta f_{gas}$ increasing with decreasing $r_{p}$ and $\left | \mu  \right |$,  whereas  $\Delta$SFE not exhibiting any trend, with only close pairs ($r_{p}$ $<$ 10 kpc) and equal-mass systems ($\left | \mu  \right |$ $\approx$ 0)  being affected significantly  (\S\ref{sec_rp}).

The results of  our work are in broad agreement with several previous  studies.
 \cite{Com94} also find a correlation of SFR and molecular gas mass (both are in raw values) with $r_{p}$. 
 At the same time, they found that the SFE shows no evidence for a correlation with $r_{p}$,  and  is only enhanced in galaxies displaying the strongest distortion, presumably analogous to the  objects with the smallest $r_{p}$  in our sample.
A similar conclusion is reached by \cite{Cas04} with  several hundred  galaxies in pairs with CO data from the literature.
They find that galaxies in pairs have more molecular gas than normal galaxies; however, the gas does not seem to be more efficient in forming stars.

Some studies propose opposite results favoring SFE as the prime driver of interaction-induced star formation.
\cite{Sol88}  find that only strong interactions show enhanced SFE.
\cite{Sof93} find an elevated SFE in Arp peculiar galaxies. 
The fact that SFE is enhanced in our close pairs is  in agreement with these studies, since these are  strongly interacting galaxies and  Arp peculiar galaxies which, by definition, must be close to each other.
However, these studies find no difference  in total gas mass  between isolated galaxies and galaxies in pairs, and  that leads to the conclusion that  SFE is the  determining factor in triggering star formation.
It is not clear where this discrepancy stems from.
The  possible sources of the discrepancy could be the choice of $\alpha_\mathrm{CO}$ and the definition of the isolated (control) galaxy sample.
In fact, many widely-separated spectroscopic pairs show no obvious  distortion in their morphology, but do show enhancement of SFR, $M_\mathrm{H_{2}}$, and $f_\mathrm{gas}$.
This emphasizes the importance of large spectroscopic data sets for identifying pairs and controls.

The  physical origin of the enhanced $M_\mathrm{H_{2}}$ and $f_{gas}$  is still unclear.
One possible reason  for the enhancement is an efficient transition from atomic  to molecular gas  by external pressure as suggested by  \cite{Kan17}.
The cause of the external pressure can be  the widespread shocks produced by interaction prevailing throughout a galaxy   and  cloud-cloud collisions in colliding regions  \citep{Ick85,Bra04,Bar04,Roc15}.
Such  an acceleration of \ion{H}{1}-to-H$_{2}$ transition occurs even in the early stages   interaction \citep{Kan17}. 
Moreover, \cite{Bra93} propose a scenario in which the gravitational torque induced by galaxy interaction provoke the infall of diffuse ionized halo gas inwards.
The ionized gas  progressively turns in to atomic gas because cooling  becomes more efficient with increasing density. 
Merger simulations by \cite{Mor18} indeed find that ionized gas is depleted during the interaction phase probed by galaxy pairs.
At a certain radius, because of the high density and cold temperature, the gas enters the molecular phase, causing a growth of the H$_{2}$ mass. 
The scenario is supported by cosmological hydrodynamical simulations \citep{Mos11} and binary merger simulations \citep{Mor18}.
Moreover, the  origin of the enhanced $M_\mathrm{H_{2}}$ and $f_{gas}$ might be somewhat analogous to what is being encountered by galaxies in clusters.
\cite{Mok16,Mok17} show that   Virgo cluster galaxies  have a significantly higher H$_{2}$ to \ion{H}{1} ratio than the group sample. 
They interpret it as being a result of the  various forms of interactions between galaxies that lead to gas flowing towards the center of host galaxies and the creation of H$_{2}$.
However, \cite{Ell15} and \cite{Ell18} have both shown that there is no decrease in \ion{H}{1} in late stage galaxy mergers.  
It could still be possible that the \ion{H}{1} reservoir fuels the H$_{2}$, but that the interplay with other phases also plays a role and replenishes the \ion{H}{1} \citep[e.g.,][]{Mor18}

The impact of  galaxy interactions is less evident for SFE (except for the closest pairs and equal-mass pairs)  when considering the globally-averaged properties.
The comparable  \emph{integrated} SFE of local early-stage pairs and isolated systems  is in line with theoretical predictions.
The simulations of \cite{Ren14}  find that, on a galaxy-wide scale, approaching pairs  are forming stars with similar efficiency as local spirals because the gravitational interaction and inflow is too weak to significantly increase the gas density.
As many of our galaxies in pairs are in an early-stage interaction  with their companion (i.e., two separated galaxies), the normal SFE of our galaxies in pairs is   not surprising.
Besides, we cannot exclude the possibility that galaxy interactions affect SFE at a much smaller scale than $f_{gas}$, such as the  nuclear region, collision front, or particular side (e.g., leading or trailing) with respect to the interaction, and thus the variation in SFE is averaged out in a galaxy-wide study. 
The small-scale variation in SFE may also contribute to the large scatter of SFE among the galaxies in pairs (Figure \ref{fig_grad_all_xco}), as the measured global SFE would   depend on the observed area of a galaxy and various projection effects.

As the merger proceeds,  the \emph{nuclear} gas surface density of gas gets boosted by gravitational torques and inflows, making the gas more efficient at converting gas into stars \citep{Mih96,Ren14,Spa16,Bus17}, and the system  moves to its starburst phase.
In agreement with  the simulations, the SFE appears to be enhanced  in our close pairs undergoing a strong interaction.
However, it is  worth cautioning that while  nuclear starbursts are  frequent among mergers (in both simulations and observations), observationally there is a significant fraction of systems in which interaction-triggered star formation is taking place outside the nuclear region (\citealp{Gar09,Cor17,Tho18}; Pan et al. in prep.).
Simulations based on the  standard star formation model, i.e., in which the local SFR is related to the local gas density,  often fail to reproduce this large-scale star formation \citep[e.g.,][]{Bar04,Chi10,Bou11} and  underestimate the SFR in  regions where gas exhibits large velocity dispersions \citep[e.g.,][]{Mih93,Bar04,Chi10,Bou11}.
Shock-induced star formation and clustered star formation  have been  suggested to  better account for the large-scale star formation in many interacting galaxies \citep[e.g.,][]{Jog92,Bar04,Sai09,Tey10,Pow13}.

Finally, we should note that, as pointed out by \cite{Vio18}, the decrease of  SFR-matched $\Delta M_\mathrm{H_{2}}$ and $\Delta f_{gas}$ implies that  internal mechanisms in isolated galaxies can have an effect similar to that caused by galaxy interactions.
Mechanisms such as  a bar instability could be a potential driver to accelerate  atomic  to molecular gas transitions \citep{Mas12} and promote star formation \citep{Mar97,Ell11,Car16,Kim17} in isolated galaxies.
However, the formation of bars may be  closely tied with galaxy interactions  \citep{No87,Lan14,Lok16}.

\subsection{SFR-$M_\mathrm{H_{2}}$ Relation}
\label{sec_sfl}
 The SFR (or SFR surface density) is observed to correlate with  $M_\mathrm{H_{2}}$ (or $M_\mathrm{H_{2}}$ surface density) with a power-law index of  $N$ $\approx$ 1 -- 2 \citep{Ken12}.
Figure \ref{fig_KS} illustrates theintegrated SFR-$M_\mathrm{H_{2}}$ relation of our galaxies in pairs (red circles) and controls (gray squares).
The two populations largely overlap on the  SFR-$M_\mathrm{H_{2}}$ plane, as also  observed for the  nearby galaxies in pairs in \cite{Vio18} (see their Figure 8).
This is  a  consequence of the similar SFE between galaxies in pairs and controls (at all but the smallest separations).
The linear least squares fits (log(SFR) $=$ $N$ $\times$ $\log M_\mathrm{H_{2}}$ $+$ $b$)  yield a  slope of $\sim$ $N$ $\approx$ 1.11 $\pm$ 0.15 for both galaxies in pairs and controls.

In order to better understand the role of galaxy interactions on star formation and  connect the different galaxy populations, we compare our galaxies with 23 local  isolated normal galaxies (green diamonds), 110 local galaxies in the Virgo clusters and nearby clouds (yellow pentagons), 19 local isolated (U)LIRGs (blue triangles), 49 local (U)LIRG mergers (orange hexagons), and 26 high-$z$ (U)LIRGs (purple thin diamonds).

\textbf{\textit{Local isolated normal galaxies and  isolated  (U)LIRGs.}} The local isolated normal galaxies and local isolated  (U)LIRGs are taken from \cite{Gao04}.
An infrared luminosity of 10$^{11}$ $L_{\sun}$ is used   to distinguish between normal galaxies and (U)LIRGs.
Only  galaxies that were observed in $^{12}$CO(1--0) with the IRAM 30-m and NRAO 12-m telescopes are used in this work.
Galaxies in \cite{Gao04}  which have been classified as a galaxy pair and a group galaxy  from the NASA/IPAC Extragalactic Database (NED)\footnote{https://ned.ipac.caltech.edu/} were removed (see Table 1 in \cite{Gao04});  however, we cannot completely rule out the possibility of these (U)LIRGs  being mergers.
Most of the normal isolated objects are  NGC galaxies, while the isolated (U)LIRGs are IRAS and Markarian galaxies.

\textbf{\textit{Local galaxies in the Virgo cluster and nearby clouds.}}
The local galaxies in dense environments, including the Virgo cluster and nearby clouds,  are taken from the Herschel Reference Sample\footnote{https://hedam.lam.fr/HRS/} \citep[HRS;][]{Bos10}.
The nearby clouds are Leo, Ursa Major and Ursa Major Southern Spur, Crater, Coma I, Canes Venatici Spur and Canes Venatici-Camelopardalis and Virgo-Libra Clouds.
Galaxies are removed from the analysis  if they have been identified as early type by NED or they are located at the Virgo outskirts.
The  molecular gas data in $^{12}$CO(1--0) for the HRS are either obtained  at the Kitt Peak 12-m telescope or compiled from the literature \citep{Bos14a}.

\textbf{\textit{Local (U)LIRG mergers.}} The measurements of local (U)LIRG mergers are compiled from \cite{Gao99}.
The observations were made with the IRAM 30-m,  NRAO 12-m, and SEST 15-m telescopes at $^{12}$CO(1--0).
The (U)LIRG mergers are IRAS, Markarian, and Arp galaxies.

\textbf{\textit{High-$z$ (U)LIRG.}} The high-$z$ (0.2 $<$ $z$ $<$ 1.0) (U)LIRGs are taken from  \cite{Com13}.
Data of multiple CO transitions were taken, but in Figure \ref{fig_KS} we only use the galaxies with $^{12}$CO(2--1) observed with the IRAM 30-m telescope, which is the lowest transition in  \cite{Com13}.
\cite{Com13} assume a  line ratio  of: 1  between the $^{12}$CO(2--1) and $^{12}$CO(1--0),  as  expected  for    warm optically thick, and thermally excited gas in starburst objects.
About half of these high-$z$ (U)LIRGs are interacting or merging systems, while  the remaining objects appear unperturbed.
Because of the increasing gas fraction with redshift, high-$z$ galaxies can easily be (U)LIRGs without violent interactions \citep[e.g.,][]{Dav10,Kar12}. 
Moreover, the high fraction of unperturbed high-$z$ (U)LIRGs is in part due to the low resolution and sensitivity  when  imaging  high-$z$ galaxies.
In those cases where the line is not detected, the upper limits of $L_\mathrm{CO}$ and $M_\mathrm{H_{2}}$  are computed at 3$\sigma$.

Due to the lack of metallicity and $M_{\ast}$ measurements necessary to calculate the physically-motivated $\alpha_\mathrm{CO}$, we apply two values for $\alpha_\mathrm{CO}$, 3.2 and 0.8, for all galaxies compiled from the literature. 
The higher $\alpha_\mathrm{CO}$ is the median value of our  galaxies in pairs and controls; the lower  $\alpha_\mathrm{CO}$ is the commonly adopted conversion factor for (U)LIRGs and distant galaxies \citep[e.g.,][]{Sol97,Dow98,Dad10,Com13}.
Both results are plotted in Figure \ref{fig_KS} with large symbols for $\alpha_\mathrm{CO}$ $=$ 3.2 and small symbols for $\alpha_\mathrm{CO}$ $=$ 0.8. 
The two symbols of a given galaxy are connected with a horizontal line, indicating the most plausible range of $M_\mathrm{H_{2}}$ for the galaxy.
The range of $M_\mathrm{H_{2}}$ of  non-detected galaxies are also computed based on the upper limits of  $L_\mathrm{CO}$.
These galaxies are indicated by a horizontal arrow.

The SFRs of the local isolated galaxies and local and high-$z$ (U)LIRGs  are calculated using infrared luminosity ($L_{\mathrm{IR}}$) \citep{Ken98b}:
\begin{equation}
\frac{\mathrm{SFR}}{\mathrm{[M_{\odot }\, yr^{-1}]}}=4.5\times 10^{-44}\frac{L_{\mathrm{IR}}}{[\mathrm{erg\, s^{-1}}]},
\label{eq_sfr_kenn}
\end{equation}
{assuming a Salpeter IMF. The SFRs of the  HSR galaxies are determined by the mean values of different SFR estimates using H$\alpha$, 24$\mu$m, FUV, and radio, along with a Salpeter IMF as well \citep{Bos15}. 
It is necessary to multiply these Salpeter SFRs  by 0.625 to transform from Salpeter IMF to Kroupa IMF.

With $\alpha_\mathrm{CO}$ $=$ 3.2, the local isolated galaxies and galaxies in the Virgo cluster and nearby clouds   populate the same regime in SFR-$M_\mathrm{H_{2}}$  space as our  galaxies in isolation and in pairs.
The SFE of local  galaxies could increase to up to $\sim$ 10$^{-8}$ yr$^{-1}$ if the lower conversion factor is used. 
The choice of the lower conversion factor for local  galaxies might not be realistic \citep[e.g.,][]{San13,Cor18}, but it characterizes a potential upper limit of local SFE.

No matter which $\alpha_\mathrm{CO}$ is used, the SFE  increases from local non-(U)LIRGs galaxies  to local (U)LIRGs and (U)LIRG mergers, and to high-$z$ (U)LIRGs.
The change of SFE across the galaxy populations can be better seen in Figure \ref{fig_KS}(b) where we plot SFR versus SFE of gas.
In other words, the high SFR of high-$z$ (U)LIRGs is  not only due to an enhancement of molecular gas reservoir, but also the enhanced  SFE of the gas.

The  different  SFE of high-$z$ and local  star-forming galaxies  have previously been considered as ``bimodal''  \citep{Dad10,Gen10}.
In other words,  local  star-forming galaxies  and high-$z$ starburst galaxies are  two distinct populations.
However, Figure \ref{fig_KS} shows that the gap in the relation that extends from  our local galaxies to the high-$z$  (U)LIRGs is filled by local (U)LIRGs, confirming the finding of \cite{Sai11}.

Some of our galaxies in pairs, both major and minor mergers,  may become infrared-bright galaxies between the first passage and final coalescence  during which star formation  is dramatically boosted \citep[e.g.,][in fact a few of our galaxies in pairs have SFR comparable to that of local (U)LIRGs as seen in Figure \ref{fig_KS}]{Dim08,Kar12,Ren14}.
\cite{Car15} find that around 65\% of local LIRGs are minor mergers (see also \citealp{Ell13} Figure 10).
However, minor mergers typically do not induce enough gas into the nuclear region to generate  ULIRG-level luminosities. 
In the local Universe, the majority of ULIRGs are triggered by almost equal-mass,  gas rich systems in advanced merger stages \citep{Das06,Hwa10,Car15}.

\begin{figure*}%
	\centering
	\includegraphics[width=0.95\textwidth]{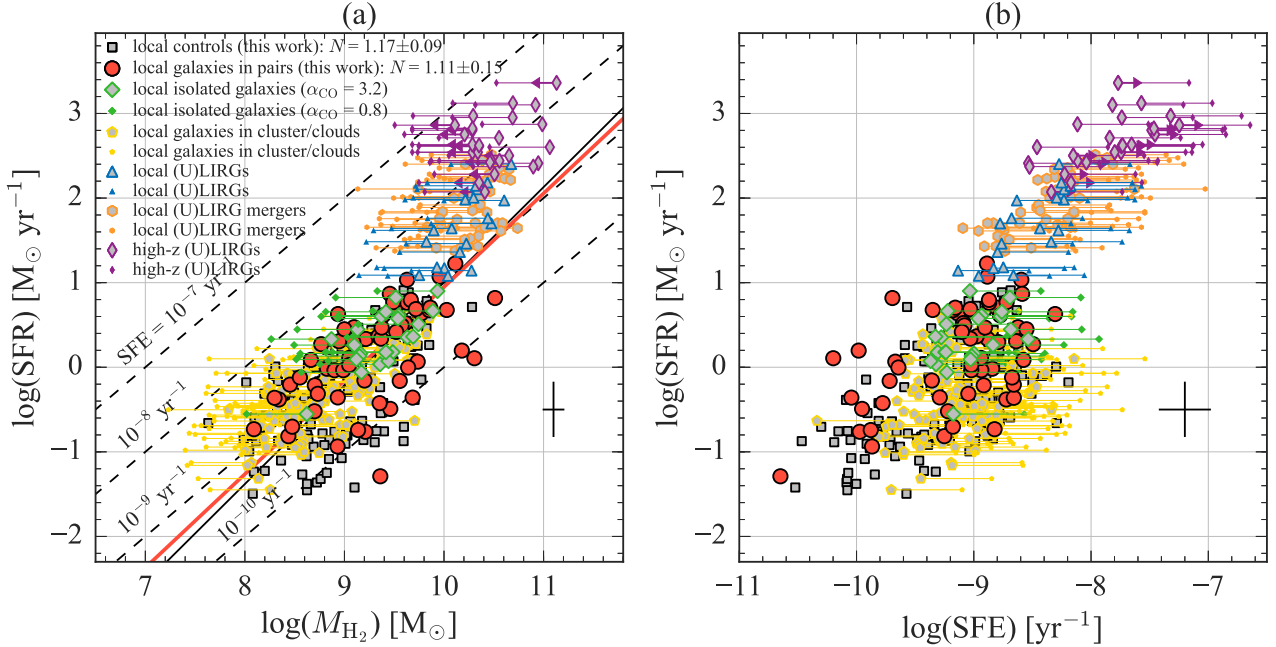}
	\caption{  SFR plotted as a function of mass (a) and star formation efficiency (b) of molecular gas. Our galaxies in pairs and the pool of controls  are shown as red circles and gray squares, respectively. Red and black solid lines  give the best-fitting linear relation for our galaxies in pairs and controls, respectively. The values of the best-fitting power law index are given in the plot.  Literature data  have been included for comparison. The local normal isolated galaxies (green diamonds) and (U)LIRGs (blue triangles) are taken from \cite{Gao04}.
	Galaxies in the Virgo cluster and nearby clouds are taken from the Herschel Reference Survey \citep[HSR, yellow pentagons; ][]{Bos10}.
	 Orange hexagons show local  (U)LIRG mergers from \cite{Gao99}. Purple thin diamonds show high-$z$ (U)LIRGs from \cite{Com13}.  Due to the lack of metallicity and $M_{\ast}$ measurements to calculate the physically-motivated $\alpha_\mathrm{CO}$, we apply two $\alpha_\mathrm{CO}$, 3.2 (large symbols) and 0.8 (small symbols), for all galaxies compiled from  literature (see text for details). The two symbols of a given galaxy are connected with a horizontal line, indicating the most plausible range of $M_\mathrm{H_{2}}$ for the galaxy.
	In the cases where the line is not detected, the upper limits of $L_\mathrm{CO}$ and $M_\mathrm{H_{2}}$  are computed at 3$\sigma$. These galaxies are indicated by  a horizontal  arrow  (all of them are high-$z$ (U)LIRGs). 
	The SFRs of the local isolated galaxies and local and high-$z$ (U)LIRGs  are calculated using $L_\mathrm{IR}$ calibrated by \cite{Ken98b}.  The SFRs of the HSR galaxies are determined by the mean values of different SFR estimates  using H$\alpha$, 24$\mu$m, FUV, and radio  \citep{Bos15}.
	The figure shows that the gap between our galaxies and high-$z$ (U)LIRGs on the  SFR-$M_\mathrm{H_{2}}$ plane (the bimodal star formation mode) can be bridged by  local (U)LIRGs. Moreover, the high SFR of high-$z$ (U)LIRGs is  not only due to an enhancement of molecular gas reservoir, but also the SFE of the molecular gas.}%
	\label{fig_KS}%
\end{figure*}

\subsection{Caveats of This Work}
\label{sec_caveats}

One caveat of the total gas mass determination is the fact that traditionally the SDSS only probes the central 3$\arcsec$. 
However, metallicity gradients  have been observed in galaxies, with a typical gradient of $\sim$ -0.05 dex kpc$^{-1}$ \citep{Pil14,Ho15,Bel17}. 
The influence of using nuclear metallicity, instead of globally-averaged metallicity, to calculate  $\alpha_\mathrm{CO}$ is presumably stronger for isolated galaxies than galaxies in pairs due to interaction-induced radial mixing of gas, which  flattens the metallicity profiles \citep{Mic08,Eli08,Kew10,Scu12,Tho18}.

Another caveat to the total gas mass determination is the assumption of $^{12}$CO(2-1) to $^{12}$CO(1-0) ratio ($R_{21}$) for the sample observed in the $J$ $=$ 2 $\rightarrow$ 1 transition.
In this work we adopted an $R_{21}$ of 0.8, which is  an average value of spatially-resolved $R_{21}$ of nearby galaxies \citep{Ler09} and corresponds to optically thick gas with an excitation temperature of $\sim$ 10 K.
However, $R_{21}$  varies from region to region in the Milky Way and in nearby galaxies: $\sim$ 0.6 -- 1.0 in the spiral arms  and galactic centers (star-forming molecular clouds) and $<$ 0.6 in  the  interarm regions (dormant molecular clouds) \citep[e.g.,][]{Ler09,Kod12,Pan15}.
If a higher value of $R_{21}$ is adopted ($R_{21}$ $\approx$ 0.9 -- 1), the $M_\mathrm{H_{2}}$, as well as  the $\Delta M_\mathrm{H_{2}}$ and $\Delta f_\mathrm{gas}$ with respect to the controls (which are observed at $J$ $=$ 1 $\rightarrow$ 0) would decrease.
However, such a high value of $R_{21}$ is only observed in the nuclear regions; given that  $R_{21}$ decreases with increasing galactocentric radius \citep{Ler09},   the true value of the globally-averaged $R_{21}$ should be lower than that.
It is also unlikely that our galaxies in pairs are dominated by dormant  clouds since they are actively forming stars, therefore a globally-averaged $R_{21}$ $<$ 0.6 is not possible.
Accordingly, we argue that  our results should be only minimally affected by the assumption for $R_{21}$. 
Nonetheless, it is important to note that a variation of $^{12}$CO(3-2) to $^{12}$CO(1-0) ratio as a function of merger sequence  of (U)LIRGs have been reported \citep{Lee10,Mic16}.

\section{Summary}
 \label{sec_summary}

We investigate the effect of galaxy interactions on global molecular gas properties by studying a sample of 58 galaxies in pairs (\S\ref{sec_mo_pairs}) and 154 control galaxies (\S\ref{sec_control_sample}).
Molecular gas properties are determined from observations with the  JCMT 15-m, PMO 14-m, CSO 10-m telescopes, and supplemented with data from  xCOLD GASS and JINGLE surveys at $^{12}$CO(2--1) and $^{12}$CO(1--0).
The main conclusions are summarized below.

\begin{enumerate}
	\item The median value of the SFR, $M_\mathrm{H_{2}}$ and $f_{gas}$ distributions of the full pairs sample are higher compared with the full control (non-merger)  sample. 
	The differences between control sample and galaxies in pairs are confirmed statistically by the  Kolmogorov-Smirnov test.
	On the other hand,  the SFE distribution of galaxies in pairs is  statistically indistinguishable from that of the control sample (\S \ref{result_absolute_prop} and Figure \ref{fig_hists}).
	
	\item We compute offsets in $M_\mathrm{H_{2}}$, $f_{gas}$ and SFR on a galaxy-by-galaxy basis by identifying  controls that are matched in redshift, stellar mass and effective size.
	All gas property offsets ($\Delta M_\mathrm{H_{2}}$, $\Delta f_{gas}$, and $\Delta$SFE)  increase with $\Delta$SFR, implying that both the available gas reservoir and SFE of the gas are expected to influence  SFR.	However, the correlations are stronger for $\Delta M_\mathrm{H_{2}}$ and $\Delta f_{gas}$ than  $\Delta$SFE in terms of  correlation coefficients (\S \ref{sec_offset}, Figure \ref{fig_hists}, and Figure  \ref{fig_scatter2_all_xcoo}).

	\item 	 $\Delta$SFR, $\Delta M_\mathrm{H_{2}}$, and $\Delta f_{gas}$ all increase with decreasing pair separation ($r_{p}$) over the range from $\sim$ 70  to 10 kpc. However, any  SFE enhancement is only significant at the smallest pair separations ($r_{p}$ $<$ 20 kpc) (\S \ref{sec_rp} and Figure \ref{fig_grad_all_xco}).
	
	\item  $\Delta$SFR, $\Delta M_\mathrm{H_{2}}$, and $\Delta f_{gas}$ also exhibit a  trend  with stellar mass ratio of the two galaxies in a pair. Statistically, higher enhancements are found in pairs with smaller mass ratio.  We find no apparent trend between the mass ratio  and  $\Delta$SFE; it seems that statistically SFE enhancements only occur  in the equal-mass pairs ($\left | \mu  \right |$ $\approx$ 0) (\S \ref{sec_mr} and Figure \ref{fig_massratio_all_xco}). 
	
	\item If we additionally match the galaxies in pairs in SFR with controls, the gas mass and fraction are still enhanced in galaxies in pairs with respect to the controls,  although  by a smaller factor (\S \ref{sec_real} and Figure \ref{fig_hist_delta_gas_prop_strmatch_xco}).

	\item Our local galaxies in pairs and controls are largely overlapping on the  SFR-$M_\mathrm{H_{2}}$ relation, as a result of their comparable SFE ($=$ SFR/$M_\mathrm{H_{2}}$). The SFE of our galaxies is   an order of magnitude lower than that in the high-$z$ (U)LIRGs. The gap between our galaxies and high-$z$ (U)LIRGs on the  SFR-$M_\mathrm{H_{2}}$ plane can be bridged by  local (U)LIRGs. Moreover,  the high SFR of high-$z$ (U)LIRGs is  not only due to an enhancement of molecular gas reservoir, but also the SFE of the gas (\S \ref{sec_sfl} and Figure \ref{fig_KS}).
		
\end{enumerate}
All the above, taken together, leads to the conclusion that  galaxy interactions do  modify the total molecular gas mass, molecular gas mass fraction, and star formation rate of a galaxy, although the strength of the effect is dependent on merger properties.

Here, we have only accounted for \emph{integrated} properties.
The next step of this work is to probe the  spatially-resolved star formation and molecular gas properties.
A direct comparison of spatially-resolved $\Delta$SFR and molecular gas properties  will  extend our understanding of the star formation process in galaxy pairs, e.g., where  the enhanced $M_\mathrm{H_{2}}$, $f_\mathrm{gas}$, and SFR are actually to be found.
ALMA observations of MaNGA galaxies will be ideal for carrying out such an analysis.

\section*{Acknowledgement}
We thank the anonymous referee for helpful suggestions that improved the paper. 
The work is supported by the Academia Sinica under the Career Development Award CDA-107-M03 and the Ministry of Science \& Technology of Taiwan under the grant MOST 107-2119-M-001-024-.
C.D.W. acknowledges support from the Natural Sciences and Engineering Research Council of Canada.
YG acknowledges support of National Key R\&D Program of China \# 2017YFA0402704, NSFC \# 11420101002 and CAS Key Research Program of Frontier Sciences.
LCH was supported by the National Key R\&D Program of China (2016YFA0400702) and the National Science Foundation of China (11473002, 11721303).
EB acknowledges support from the UK Science and Technology Facilities Council [grant number ST/M001008/1].
M.J.M.~acknowledges the support of
the National Science Centre, Poland, through the POLONEZ grant
2015/19/P/ST9/04010;
this project has received funding from the European Union's Horizon
2020 research and innovation programme under the Marie
Sk{\l}odowska-Curie grant agreement No. 665778.

The authors wish to recognize and acknowledge the very significant cultural role and reverence that the summit of Maunakea has always had within the indigenous Hawaiian community.  We are most fortunate to have the opportunity to conduct observations from this mountain.

We are grateful to the MPA-JHU group for access to their data products and catalogs. The Starlink software \citep{Cur14} is currently supported by the East Asian Observatory.

The James Clerk Maxwell Telescope is operated by the East Asian Observatory on behalf of The National Astronomical Observatory of Japan; Academia Sinica Institute of Astronomy and Astrophysics; the Korea Astronomy and Space Science Institute; the Operation, Maintenance and Upgrading Fund for Astronomical Telescopes and Facility Instruments, budgeted from the Ministry of Finance (MOF) of China and administrated by the Chinese Academy of Sciences (CAS), as well as the National Key R\&D Program of China (No. 2017YFA0402700). Additional funding support is provided by the Science and Technology Facilities Council of the United Kingdom and participating universities in the United Kingdom and Canada.
This material is based upon work at the Caltech Submillimeter Observatory, which is operated by the California Institute of Technology.

This research has made use of data from HRS project. HRS is a Herschel Key Programme utilising Guaranteed Time from the SPIRE instrument team, ESAC scientists and a mission scientist.
The HRS data was accessed through the Herschel Database in Marseille (HeDaM - http://hedam.lam.fr) operated by CeSAM and hosted by the Laboratoire d'Astrophysique de Marseille.

This research has made use of the NASA/IPAC Extragalactic Database (NED), which is operated by the Jet Propulsion Laboratory, California Institute of Technology, under contract with the National Aeronautics and Space Administration.

Funding for the Sloan Digital Sky Survey IV has been provided by the Alfred P. Sloan Foundation, the U.S. Department of Energy Office of Science, and the Participating Institutions. SDSS acknowledges support and resources from the Center for High-Performance Computing at the University of Utah. The SDSS web site is www.sdss.org.

SDSS is managed by the Astrophysical Research Consortium for the Participating Institutions of the SDSS Collaboration including the Brazilian Participation Group, the Carnegie Institution for Science, Carnegie Mellon University, the Chilean Participation Group, the French Participation Group, Harvard-Smithsonian Center for Astrophysics, Instituto de Astrofísica de Canarias, The Johns Hopkins University, Kavli Institute for the Physics and Mathematics of the Universe (IPMU) / University of Tokyo, the Korean Participation Group, Lawrence Berkeley National Laboratory, Leibniz Institut für Astrophysik Potsdam (AIP), Max-Planck-Institut f\"ur Astronomie (MPIA Heidelberg), Max-Planck-Institut f\"ur Astrophysik (MPA Garching), Max-Planck-Institut f\"ur Extraterrestrische Physik (MPE), National Astronomical Observatories of China, New Mexico State University, New York University, University of Notre Dame,  Observat\'orio Nacional / MCTI, The Ohio State University, Pennsylvania State University, Shanghai Astronomical Observatory, United Kingdom Participation Group, Universidad Nacional  Aut\'onoma de  M\'exico, University of Arizona, University of Colorado Boulder, University of Oxford, University of Portsmouth, University of Utah, University of Virginia, University of Washington, University of Wisconsin, Vanderbilt University, and Yale University.

\end{document}